\documentclass[%
 superscriptaddress,
 reprint,
 showpacs,preprintnumbers,
 nofootinbib,
 amsmath,amssymb,
 aps,
 prd,
 floatfix,
]{revtex4-1}
\usepackage{graphicx}
\usepackage{xcolor}
\usepackage{hyperref}
\hypersetup{
    colorlinks=true,
    linkcolor=blue,
    filecolor=blue,      
    urlcolor=blue,
    pdftitle={Uncertainties in SF model for LBL OA}, 
    }
\usepackage{lipsum}
\usepackage{amssymb,amsmath,amstext,amsthm,amsfonts}
\usepackage[utf8]{inputenc}
\usepackage{float}
\usepackage{url}
\usepackage{soul}
\usepackage{esvect}
\usepackage{multirow}
\usepackage{enumitem}
\usepackage[export]{adjustbox}

\RequirePackage{xspace}

\newcommand{\pt}{\ensuremath{p_\textrm{T}}\xspace}

\newcommand{\dalphat}{\ensuremath{\delta\alpha_\textrm{T}}\xspace}

\newcommand{\dpt}{\ensuremath{\delta \pt}\xspace}

\newcommand{\rvline}{\hspace*{-\arraycolsep}\vline\hspace*{-\arraycolsep}}

\newcommand{\minerva}{MINERvA\xspace}

\newcommand{\etal}{\textit{et al.}\xspace}

\usepackage{lineno}

\definecolor{LightCyan}{rgb}{0.88,1,1}
\usepackage{xcolor,colortbl}

\usepackage{todonotes} 

\begin{document}

\title{Parametrized uncertainties in the spectral function model of neutrino charged-current quasielastic interactions for oscillation analyses}

%
\author{J.~Chakrani}
\email[Contact e-mail: ]{jaafar.chakrani@polytechnique.edu}
\affiliation{Laboratoire Leprince-Ringuet, CNRS, Ecole polytechnique, Institut Polytechnique de Paris, Palaiseau, France}
	
\author{S.~Dolan}
\email[Contact e-mail: ]{stephen.joseph.dolan@cern.ch}
\affiliation{European Organization for Nuclear Research (CERN), Geneva, Switzerland}

\author{M.~Buizza Avanzini}
\email[Contact e-mail: ]{buizza@llr.in2p3.fr}
\affiliation{Laboratoire Leprince-Ringuet, CNRS, Ecole polytechnique, Institut Polytechnique de Paris, Palaiseau, France}

\author{A.~Ershova}
\affiliation{IRFU, CEA, Université Paris-Saclay, Gif-sur-Yvette, France}

\author{L.~Koch}
\affiliation{Institut f\"ur Physik, Johannes Gutenberg-Universit\"at, Mainz, Germany}

\author{K.~McFarland}
\affiliation{University of Rochester, Department of Physics and Astronomy, Rochester, New York, U.S.A}

\author{G.~D.~Megias}
\affiliation{Departamento de Física Atómica, Molecular y Nuclear, Universidad de Sevilla, 41080 Sevilla, Spain}

\author{L.~Munteanu}
\affiliation{European Organization for Nuclear Research (CERN), Geneva, Switzerland}

\author{L.~Pickering}
\affiliation{Royal Holloway University of London, Department of Physics, Egham, Surrey, United Kingdom}

\author{K.~Skwarczynski}
\affiliation{National Centre for Nuclear Research, Warsaw, Poland}

\author{V. Q. Nguyen}
\affiliation{Laboratoire Leprince-Ringuet, CNRS, Ecole polytechnique, Institut Polytechnique de Paris, Palaiseau, France}

\author{C.~Wret}
\affiliation{Oxford University, Oxford, United Kingdom}


\begin{abstract}

\noindent 
A substantial fraction of systematic uncertainties in neutrino oscillation experiments stem from the lack of precision in modeling the nuclear target in neutrino-nucleus interactions.
Whilst this has driven significant progress in the development of improved nuclear models for neutrino scattering, it is crucial that the models used in neutrino data analyses be accompanied by parameters and associated uncertainties that allow the coverage of plausible nuclear physics. Based on constraints from electron scattering data, we develop such a set of parameters, which can be applied to nuclear shell models, and test their application to the Benhar \etal spectral function model. The parametrization is validated through a series of maximum likelihood fits to cross-section measurements made by the T2K and MINERvA experiments, which also permit an exploration of the power of near-detector data to provide constraints on the parameters in neutrino oscillation analyses.
\end{abstract}

\maketitle

\section{Introduction}
\label{sec:introduction}
Modeling neutrino-nucleus interactions remains a major challenge for current and future neutrino oscillation experiments~\cite{Alvarez-Ruso:2017oui}. Whilst statistical uncertainties remain large in the ongoing accelerator-based long-baseline neutrino oscillation experiments, T2K~\cite{T2K:2019bcf} and NO$\nu$A~\cite{NOvA:2018gge}, the future experiments, Hyper-Kamiokande (HK)~\cite{Hyper-Kamiokande:2018ofw} and DUNE~\cite{DUNE:2020lwj}, are likely to be dominated by uncertainties related to neutrino interactions. To tackle this issue, experiments are designing and constructing more sophisticated near detectors~\cite{T2K:2019bbb,DUNE:2021tad,Hyper-Kamiokande:2018ofw}. However, more accurate neutrino interaction models are a critical ingredient to precisely extrapolating new near-detector measurements into robust constraints on the uncertainties in neutrino oscillation analyses. Beyond just incorporating better models, it is also necessary for future neutrino oscillation analyses to include a comprehensive set of uncertainties to cover their plausible variations. Establishing such lists of uncertainties remains one of the most significant challenges for robustly reducing systematic uncertainties using near-detector data in neutrino oscillation analyses. 

At $E_\nu\sim 1~\textrm{GeV}$ energy, neutrino interactions with nucleons bound within nuclear targets are significantly impacted by nuclear effects which can have a substantial impact on the interaction cross section and on the kinematics of the final-state particles. These can affect the bias in metrics for reconstructing neutrino energy from observable final-state interaction products which, if modeled incorrectly, directly biases the measurement of neutrino oscillation parameters~\cite{Alvarez-Ruso:2017oui}. 
In particular, due to the removal energy and the Fermi motion of the bound nucleon, the neutrino interacts on an off-shell, non-static particle which can significantly alter interaction kinematics with respect to the case of a free-nucleon target. Multiple processes can further affect the final-state kinematics: this is the case for Pauli blocking (PB), which prevents some interactions to occur, and for final-state interactions (FSI), which describe how the interaction between the outgoing hadrons and the residual nucleus affects the process. 

A relatively sophisticated way to model charged-current quasielastic interactions (CCQE)---the dominant interaction channel for T2K and HK---is provided by the spectral function model (SF) by Benhar \etal~\cite{Benhar:1994hw}. It is a nuclear shell model built largely from electron scattering data, and allows a detailed description of the initial nuclear state. Through its implementation in the \texttt{NEUT} neutrino-nucleus interaction event generator~\cite{Hayato:2021heg}, the SF model is used in the T2K experiment's recent measurements of neutrino oscillations~\cite{T2K:2023smv} and neutrino-nucleus cross sections~\cite{T2K:2018rnz, T2K:2020jav,T2K:2023zaf,T2K:2020sbd}. It has also been used to study the sensitivity of T2K's near-detector upgrade~\cite{Dolan:2021hbw}. 

Shell-based models lend themselves to consistently defining an ensemble of uncertainties which affect outgoing lepton and hadron kinematics in neutrino interactions. The shell structure presents natural degrees of freedom that can be varied, and their determination from electron scattering analyses can motivate possible model variations. In this paper, we detail a set of parameters to alter the SF alongside associated uncertainties. The general scheme is applicable to any shell-based model. The parameters applied to the SF model are benchmarked by fitting them to neutrino-nucleus cross-section measurements from the T2K and \minerva experiments. 

The paper is organized as follows: in Sec.~\ref{sec:SF} the Benhar \etal SF model is summarized, and Sec.~\ref{sec:syst} introduces a parametrization of systematic uncertainties for the SF model as well as for other physics processes that affect comparisons with cross-section measurements.
In Sec.~\ref{sec:fit} we show how the use of this parametrization in fits of the \texttt{NEUT} generator to existing neutrino cross-section measurements can improve the agreement of the SF model with recent measurements from the T2K and \minerva experiments. Means of mitigating effects from ``Peelle's pertinent puzzle'' (PPP)~\cite{DAgostini:1993arp,ppp} when fitting the measurements are also presented and applied. In Sec.~\ref{sec:impactOnOA} a qualitative discussion of the parametrization's impact on neutrino oscillation analyses is presented, alongside an assessment of the scope for possible constraints from near detectors. Finally, the discussion and conclusions are presented in Sec.~\ref{sec:concl}.





\section{The Benhar spectral function model for neutrino oscillation experiments}
\label{sec:SF}



\begin{figure}
    \centering
    \includegraphics[width=\linewidth]{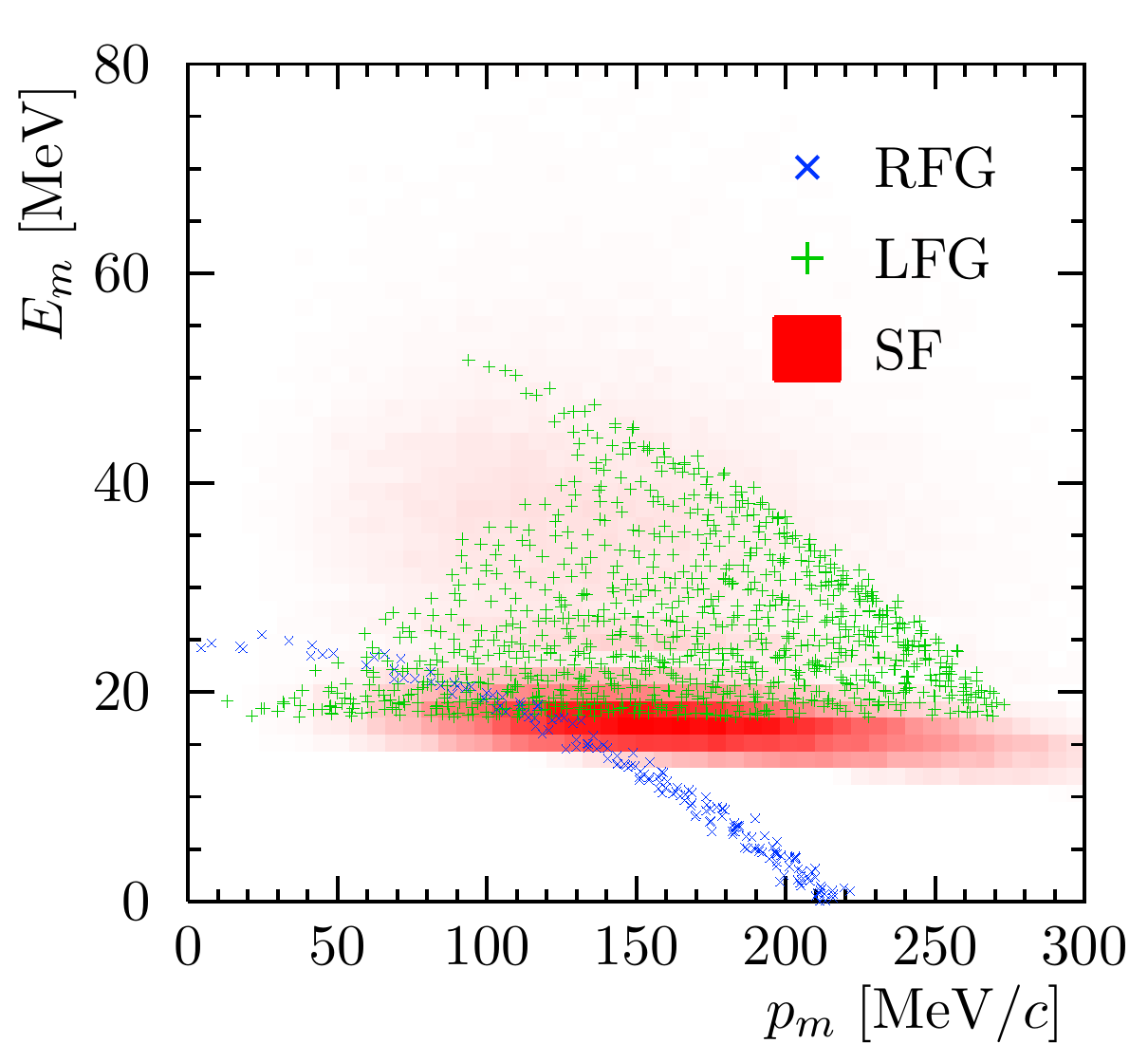}
    \caption{Simulations of the two-dimensional distribution of missing energy, $E_m$, and missing momentum, $p_m$, for carbon using the RFG (blue), LFG (green) and SF (red) model simulated with the \texttt{NEUT} neutrino interaction generator.}
    \label{fig:2DSFvsLFGvsRFG}
\end{figure}

Neutrino experiments rely on models available in Monte Carlo event generators, such as \texttt{NEUT}~\cite{Hayato:2021heg}, \texttt{NuWro}~\cite{Zmuda:2015twa} and \texttt{GENIE}~\cite{Andreopoulos:2009rq}, to provide a model of neutrino-nucleus interactions for neutrino oscillation measurements. 
Various prescriptions describe the initial state of the bound nucleons within a target nucleus. Generally speaking, neutrino oscillation experiments use either a relativistic global Fermi gas (RFG) model, a local Fermi gas (LFG) model, or a spectral function (SF) model. 
The RFG is a simplistic approach that describes the nuclear ground state and Fermi motion, which has been commonly-used to model GeV-scale neutrino scattering. It considers the target nucleons as independent fermions within a constant binding potential. 
The LFG model is a more realistic approach that introduces the radial dependence in the nuclear potential under the local density approximation (LDA). 
The SF model provides a sophisticated description of the nuclear ground state, featuring a shell structure of the nucleus as observed in electron scattering data, with an additional theory-driven component to describe ``short-range correlations'' (SRC) between nucleons within the nucleus. 
Fig.~\ref{fig:2DSFvsLFGvsRFG} shows a comparison of the predictions between the three models for the distribution of the nucleon removal (or \textit{missing}) energy, $E_m$, and the initial-state nucleon (or \textit{missing}) momentum, $p_m$, in carbon. Closely following the prescription in Ref.~\cite{JLabE91013:2003gdp}, for CCQE neutrino-nucleus interactions with a single nucleon in the final state before consideration of FSI (i.e.\ $\nu + A \rightarrow \ell + N + A'$, where $A$ and $A'$ are the initial-state nucleus and the final-state remnant respectively and $N$ is the outgoing nucleon), these can be defined as\footnote{Whilst \textit{removal} and \textit{missing} energy, and initial state and \textit{missing} momentum, are equivalent here, it should be noted that this is for the special case of single nucleon knock out with no FSI considered within this paper. The \textit{missing} energy and momentum observables reported by electron scattering experiments (as in Ref.~\cite{JLabE91013:2003gdp})  must usually be corrected in order to obtain removal energy and initial state momentum.}:
\begin{equation}
    E_m = E_\nu - E_\ell - T_N - T_{A'} - \Delta m_N,
\end{equation}
\begin{equation}
    p_m = |\vec p_m| = |\vec p_\nu - \vec p_\ell - \vec p_N|,
\end{equation}
where $E_\nu$ and $\vec p_\nu$ ($E_\ell$ and $\vec p_\ell$) are the incoming neutrino (outgoing charged lepton) energy and momentum respectively; $T_N$ and $p_N$ are the kinetic energy and momentum of the pre-FSI outgoing nucleon respectively; $T_{A'}$ is the kinetic energy of the nuclear remnant; and $\Delta m_N$ accounts for the mass difference of the initial- and final-state nucleon. $T_{A'}$ is calculated from $p_N$ assuming a mass of the ground state of the remnant nucleus following the interaction.

Fig.~\ref{fig:2DSFvsLFGvsRFG} clearly highlights the nuclear shell structure of carbon encoded in the SF model, with a sharp $p$-shell at $E_m \sim 18$~MeV and a diffuse $s$-shell at $E_m \sim 35$~MeV, which is not captured by the Fermi gas-based models. In Fermi-gas models the kinetic energy from the Fermi motion of the stuck nucleon contributes to overcoming the removal energy which gives the parabolic shapes observed. Both SF and LFG have a minimum removal energy which is derived from electron scattering data or the difference between initial state and remnant nucleus masses respectively (for the LFG case, more details can be found in~\cite{Bourguille:2020bvw}).

\begin{figure*}[]
    \centering
    \includegraphics[width=0.4\linewidth,valign=t]
    {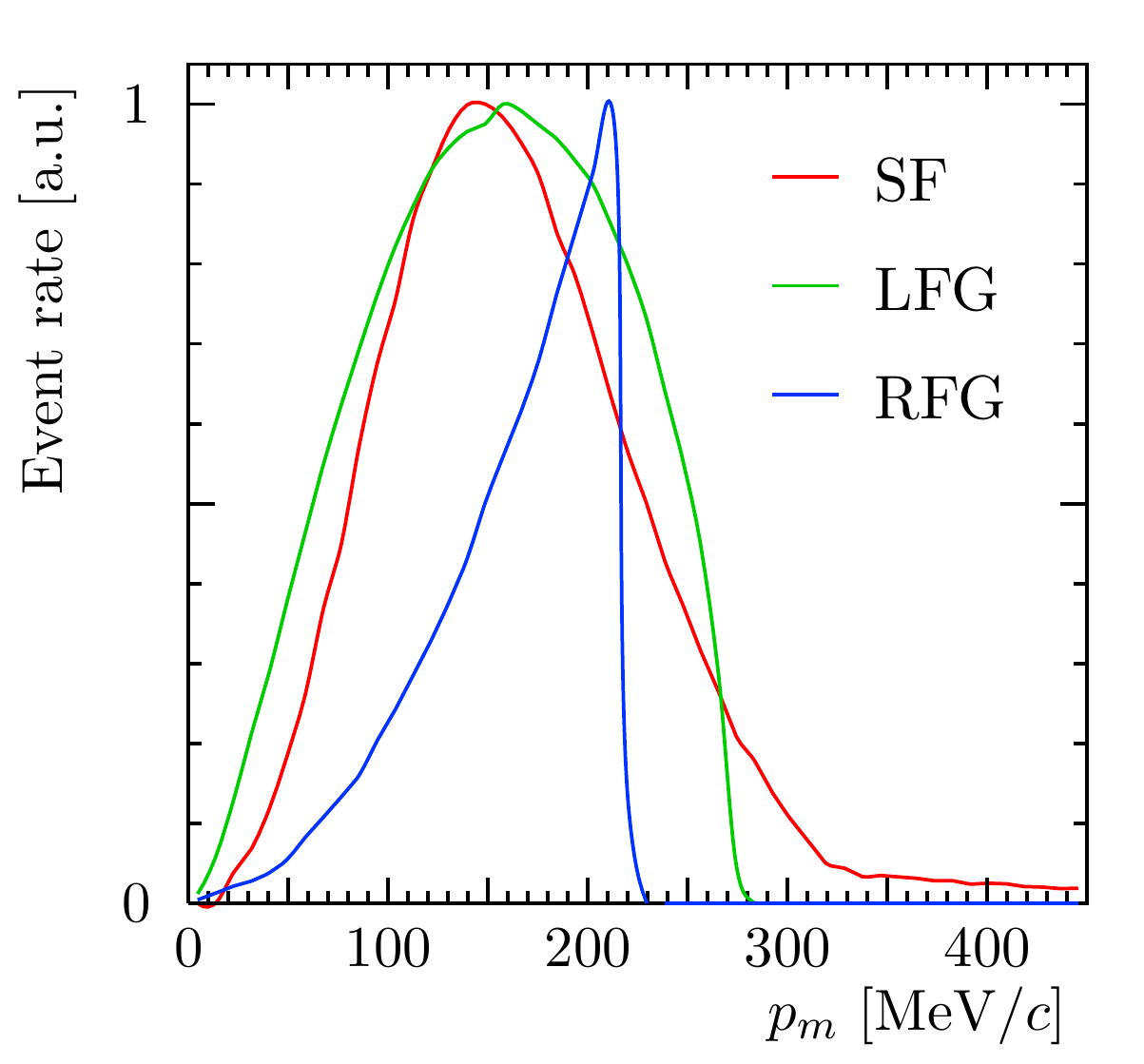}
    \includegraphics[width=0.4\linewidth,valign=t]{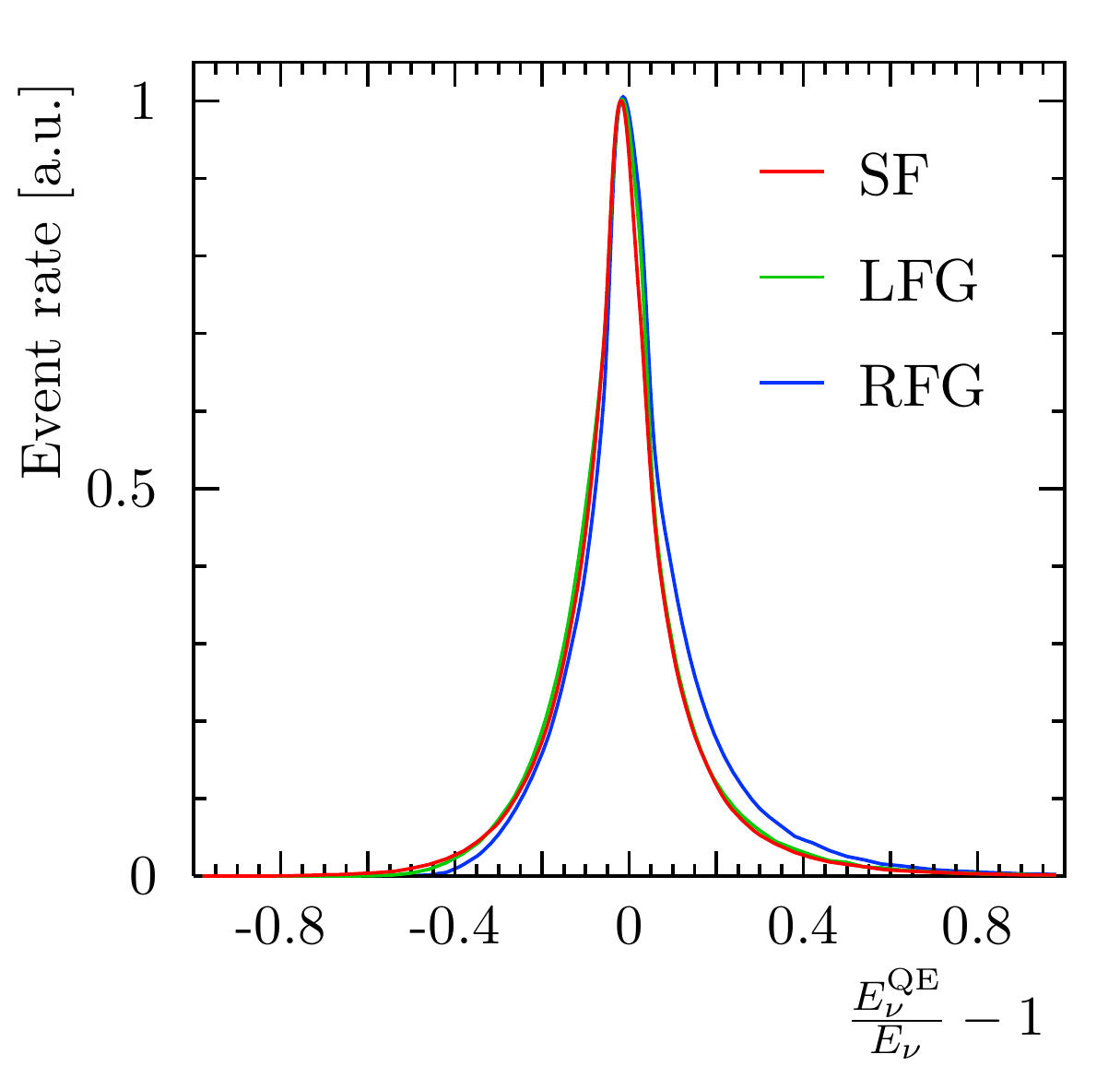}
    \caption{Comparison between the distribution of the missing momentum (left) and the neutrino energy bias (right) for CCQE interactions for the RFG (blue), LFG (green) and SF (red) model generated by \texttt{NEUT} using the T2K muon neutrino flux.}
    \label{fig:RFGvsLFGvsSF}
\end{figure*}

The distribution of the initial-state nucleon momentum is also different between the models, as displayed most clearly in the left panel of Fig.~\ref{fig:RFGvsLFGvsSF}. The RFG model assumes that $p_m$ is uniformly distributed in 3-momentum, $\vec{p}$, below the Fermi momentum $p_F$, which yields a cliff feature at $p_m\sim 220$~MeV/$c$. The LFG and SF models both predict narrower $p_m$ distributions than the RFG model.
A high momentum tail appears only in the SF model, a direct consequence of the contributions from SRCs.

These differences in nuclear ground-state modeling can manifest themselves in important ways for neutrino oscillation analyses. Of particular importance is the impact they can have in determining the bias of neutrino energy reconstruction. The oscillation analyses in T2K and HK focus on single-track events, meaning the selections are dominated by CCQE, two-particle two-hole (2p2h), and single-pion interactions where the pion is missed or absorbed. In single-track events, the incoming neutrino's energy is estimated using only the kinematics of the outgoing charged lepton by assuming that the interaction is CCQE on an initial-state nucleon at rest, with an average binding energy $E_b$, and that there is a single outgoing nucleon\footnote{In the case of single-track events intended to isolate single pion events, the outgoing nucleon mass can be replaced by the $\Delta(1232)$ mass.},
\begin{equation}
    E_\nu^\mathrm{QE} = \frac{2 E_{l} \tilde{m}_N-\left(m_l^2+\tilde{m}_N^2-m_N^2\right)}{2\left(\tilde{m}_N-E_{\ell}+p_l \cos \theta_l\right)}
\end{equation}
where $m_N$ is the mass of the struck nucleon, $\tilde{m}_N = m_N - E_b$, and $m_l$, $E_l$, $p_l$, and $\theta_l$ are the mass, the energy, the momentum, and the angle between the charged lepton and the neutrino, respectively. The right panel of Fig.~\ref{fig:RFGvsLFGvsSF} shows the bias of this estimator with respect to the true neutrino energy, $E_\nu$, for the different ground-state models. The smearing is dominated by the isotropic Fermi motion of the bound nucleon. The simplistic RFG model is offset due to its shifted $E_m$ distribution with respect to the other models, as shown in Fig.~\ref{fig:2DSFvsLFGvsRFG}, while the LFG and SF models are more similar. Since the SF model accounts for the SRC contribution, $E_\nu^\mathrm{QE}$ tends to underestimate the neutrino energy for these events, hence the longer tail in the negative region. 

\begin{figure*}[]
    \centering
    \includegraphics[width=0.4\linewidth]{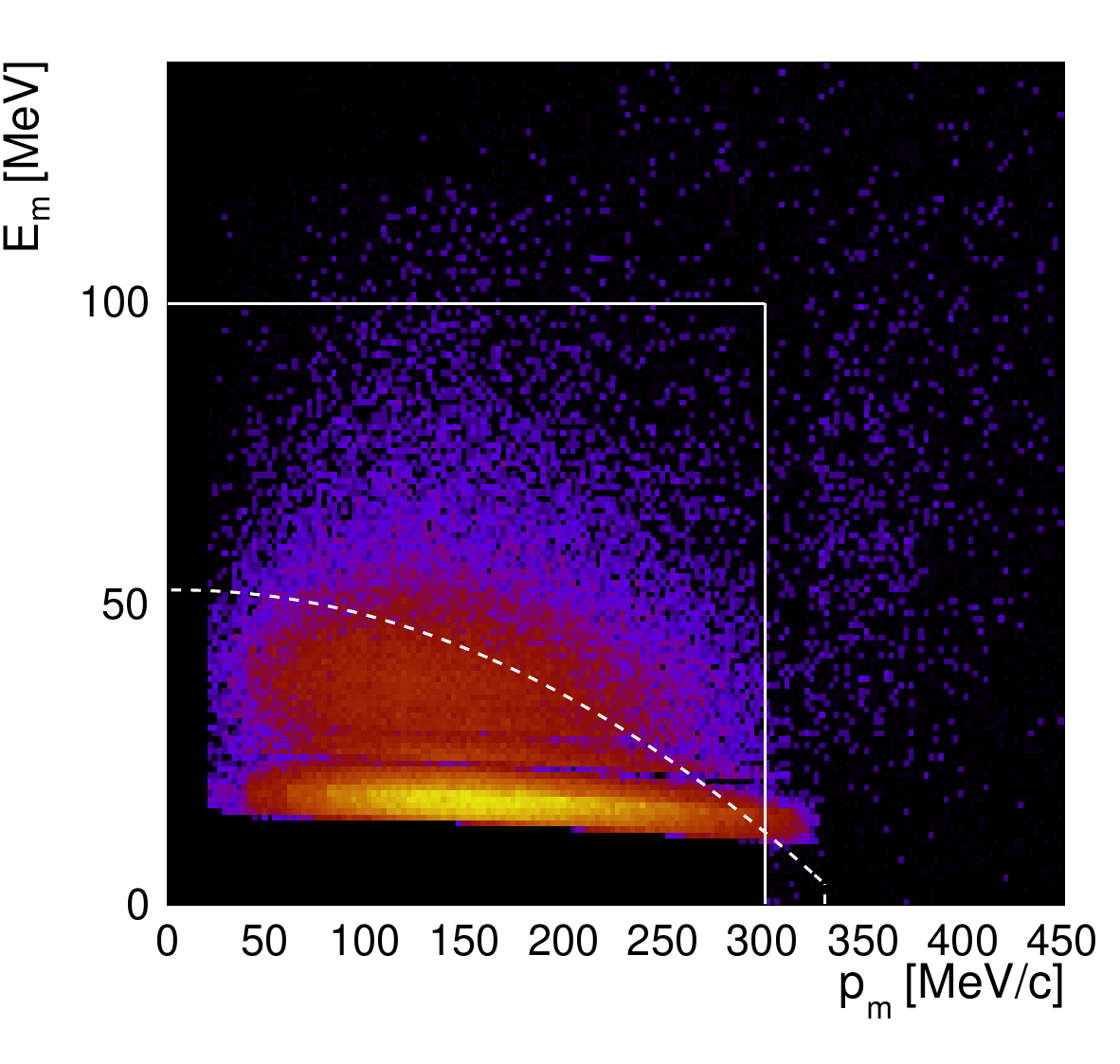}
    \includegraphics[width=0.4\linewidth]{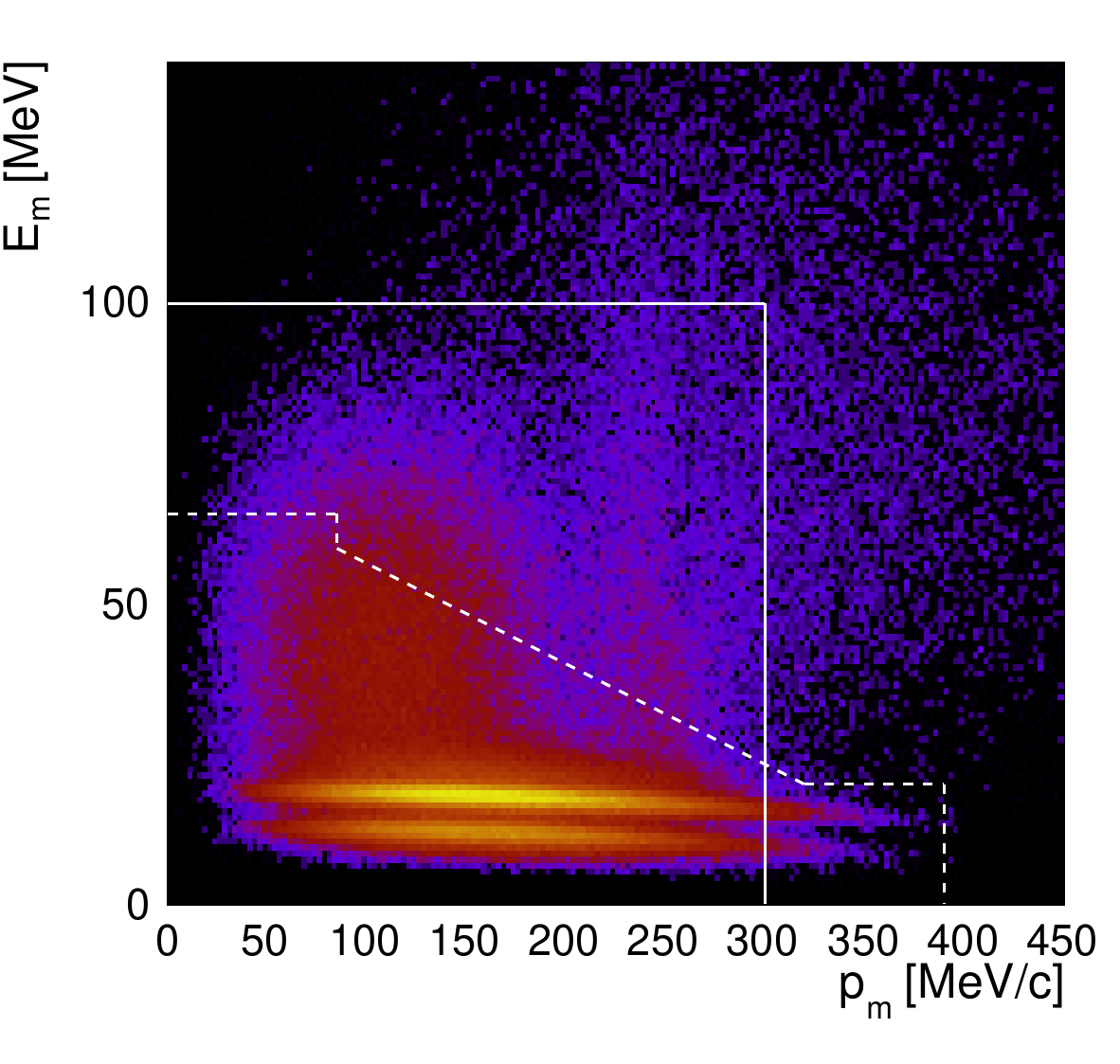}
    \caption{The prediction of the two-dimensional distribution of missing energy and missing momentum for carbon (left) and oxygen (right) for \texttt{NEUT}. The brightness of the color represents the probability of finding an initial-state nucleon with a particular removal energy and momentum state. The white lines indicate the cuts used to separate the MF region (low $E_m, p_m$) from the SRC region (high $E_m, p_m$) in the \texttt{NuWro} (dashed) and \texttt{NEUT} (solid) neutrino interaction generators.}
    \label{fig:2DSFNEUT}
\end{figure*}

The SF model, which is the focus of this work, provides a description of the two-dimensional distribution of the missing energy, $E_m$, and the missing momentum, $p_m$, of the nucleons within the nucleus. It is implemented in both the \texttt{NEUT} and \texttt{NuWro} event generators, where the joint probability distribution of $E_m$ and $\vec{p}_m$ of nucleons within the nucleus is written as:
\begin{equation}
    P(\vec{p}_m, E_m)=P_\mathrm{MF}(\vec{p}_m, E_m)+P_{\mathrm{corr}}(\vec{p}_m, E_m),
\end{equation}
where $P_\mathrm{MF}(\vec{p}_m, E_m)$ is the mean field (MF) part and $P_\mathrm{corr}(\vec{p}_m, E_m)$ corresponds to the contribution from SRC nucleons. The MF term describes the nucleus as a shell model and is tightly constrained with electron scattering data. Within neutrino event generators, SRC come from generally high $p_m$ (above the RFG Fermi level) CCQE interactions, modeled as if with a single ``target'' nucleon but where a ``spectator'' nucleon is also present. The spectator has similar magnitude but opposite direction $\vec{p}_m$ to the target nucleon. Prior to interaction, the spectator and target only remain inside the nucleus because of their correlation maintaining a collectively bound state. Once the target nucleon is ejected from the nucleus, the spectator, despite not receiving any of the interaction's momentum transfer, has too high $p_m$ to remain bound without its pair and so also leaves the nucleus. It should be noted that SRCs are implemented in generators as being distinct from 2p2h, which directly consider the cross section from a series of Feynman diagrams in which the neutrino interaction takes place with two nucleons bound by meson exchange currents and that the momentum transfer of the interaction is shared between them~\cite{Dolan:2019bxf,Schwehr:2016pvn}.
Whilst the original SF model calculates the MF and SRC components of the SF separately, the implementation in event generators is as a single integrated SF which the generators then choose to factorize into an MF and SRC parts as disjunct areas in $E_m$, $p_m$ space. 
The two-dimensional (${p}_m, E_m$) distributions for oxygen and carbon are shown in Fig.~\ref{fig:2DSFNEUT}, where the MF and SRC regions as defined in \texttt{NEUT} (solid) and \texttt{NuWro} (dashed) are also drawn. 

Although SF represents one of the more sophisticated nuclear ground-state models implemented in neutrino event generators, it is still, like most other implemented models, based mostly on the plane-wave impulse approximation (PWIA). PWIA assumes that the interaction cross section can be calculated as the incoherent sum of scattering from free nucleons with a given initial energy and momentum, and that the outgoing nucleon is not affected by the nuclear potential (i.e.\ it exits as a plane wave). In other words, only one nucleon of the target nucleus is involved in the interaction, and there is no consideration of FSI as the nucleon propagates through the nucleus at the level of the cross-section calculation.

Within the SF model's implementation in neutrino event generators, several effects beyond PWIA are considered using \textit{ad-hoc} approaches. These include Pauli blocking (PB) and FSI via an intranuclear cascade~\cite{Dytman:2021ohr} and/or a correction to the cross section based on the distortion of the outgoing nucleon via a nuclear potential (see e.g.~\cite{Gonzalez-Jimenez:2021ohu, Franco-Patino:2022tvv}). PB accounts for the statistical correlations between the struck nucleon and the remnant nucleus. A simplistic approach to estimate its effect, which is used in \texttt{NEUT}, consists of modifying the spectral function by:
\begin{equation}
P(\vec{p}_m, E_m) \rightarrow P(\vec{p}_m, E_m) \theta\left(|\vec{p}_m+\vec{q}|-{p}_F\right),
\label{eq:PB}
\end{equation}
where $\theta$ is the Heaviside step function, $\vec q$ is the transferred momentum, and $p_F$ is the Fermi surface momentum, which is nucleus dependent. This amounts to considering that all the states below Fermi momentum are occupied, which suppresses a portion of the available phase space for the outgoing nucleon. Other prescriptions are also possible, such as the LDA-based PB implemented in \texttt{NuWro}.

PB is a first-order effect beyond the impulse approximation part of PWIA. Whilst PWIA is justifiable for high energy transfer, the interactions with low energy transfer ($q_0\lesssim70$~MeV)---which constitute a significant fraction ($\sim20$\%) of CCQE neutrino interactions in oscillation experiments---need to account for other effects beyond the PWIA. In the SF model, one such effect is the aforementioned impact of FSI on the cross section. The impact of FSI on outgoing nucleon kinematics is usually included in event generators via hadron transport models such as intranuclear cascades~\cite{Dytman:2021ohr} with re-interaction probabilities derived from data. Another approach to considering the impact of FSI on an SF model is presented in Ref.~\cite{Ankowski:2014yfa}, where modifications to the inclusive cross section are calculated by considering the outgoing nucleon in a nuclear optical potential. The relationship between these two approaches to describing FSI is described in Ref.~\cite{Nikolakopoulos:2022qkq}. Within event generators, \texttt{NEUT} only implements an intranuclear cascade whilst \texttt{NuWro} offers both a cascade and an optional optical potential correction (which is not applied in Fig.~\ref{fig:2DSFNEUT}). 

Recent neutrino-nucleus cross-section measurements provide a crucial benchmark for all these models. Whilst CCQE-focused measurements which integrate over nucleon kinematics often favour Fermi gas-based models, thanks to the strong ``RPA'' correction to account for correlations between initial-state nucleons applied as a step away from the impulse approximation~\cite{Nieves:2011pp}, measurements that include nucleon kinematics tend to disfavor them and prefer the SF model due to its predictive power of the outgoing nucleon kinematics (although it should be noted that no model is able to achieve reasonable agreement with all data)~\cite{Avanzini:2021qlx,Dolan:2018zye}. 
For instance, the single-transverse variables~\cite{Lu:2015tcr} allow a probe of nuclear effects, particularly with the measurement of the transverse momentum imbalance $\dpt$ defined from the kinematics of the outgoing charged lepton and nucleon as:
\begin{equation}
    \delta \vec{p}_\mathrm{T} =  \vec{p}^\mathrm{T}_l + \vec{p}^\mathrm{T}_N,
\end{equation}
where $\vec p^\mathrm{T}_l$ and $\vec p^\mathrm{T}_N$ are the momenta of the charged lepton and the struck nucleon projected on the transverse plane with respect to the incident neutrino direction, respectively. It corresponds to the transverse projection of the Fermi motion of the initial-state nucleon, modified by nuclear effects such as FSI. Measuring \dpt therefore probes the nuclear effects experienced by the struck nucleons. The bulk of its distribution is mainly due to Fermi motion, while its tail is more sensitive to FSI and multinucleon processes. The T2K~\cite{T2K:2018rnz} and \minerva~\cite{MINERvA:2018hba} measurements of \dpt show that the SF model is indeed more suitable than Fermi-gas models to describe their measurements (although it remains unable to give a quantitatively good description of all the measurements). 
Recent analyses from MicroBooNE~\cite{MicroBooNE:2023dwo} offer new measurements, which have not yet been confronted with SF models.

\section{Systematic uncertainties}
\label{sec:syst}


As discussed in Sec.~\ref{sec:SF}, the SF model offers a more detailed description of the nuclear medium in comparison with widely-used alternative models like RFG or LFG. For this reason, the T2K experiment opts to use the SF model as a reference for CCQE interactions~\cite{T2K:2023smv}. As described in Sec.~\ref{sec:introduction}, a detailed investigation of the potential uncertainties related to the SF (or PWIA-based shell models in general) and the development of a comprehensive associated uncertainty parametrization are of great interest. Here, we present a parametrization of systematic uncertainties that are subsequently benchmarked in Sec.~\ref{sec:fit} via fits to cross-section measurements.

In addition to defining an uncertainty parametrization, prior values for the parameters' constraints are also assigned for fits in Sec.~\ref{sec:fit}---from both experimental data and theoretical arguments. In general, we opt for loose conservative priors, thereby allowing the data to dominate the constraints on most of the parameters in the fits. This approach is motivated by the fact that priors from experimental data might not be entirely applicable for neutrino scattering (e.g.\ taking priors from electron scattering data) or that the prior constraints are based on model-dependent assumptions. The exact priors used are stated and justified in the following sub-sections. Overall, Tab.~\ref{tab:priors} summarizes all the parametrized systematic uncertainties discussed in this section, along with their central values and prior uncertainties. 

\subsection{Shell-model uncertainties}

Nuclear shell models provide several natural degrees of freedom that can be parametrized and varied as systematic uncertainties. The SF model contains several discrete shells with a particular size and shape in addition to an SRC component. The occupancy and shape of each shell as well as the relative strength of the SRC component provide such natural freedoms. Moreover, the degree to which these can vary can be inspired by the electron scattering data that informs the nominal SF contribution.


\subsubsection{Shell occupancy}

\begin{figure}[]
    \centering
    \includegraphics[width=0.95\linewidth]{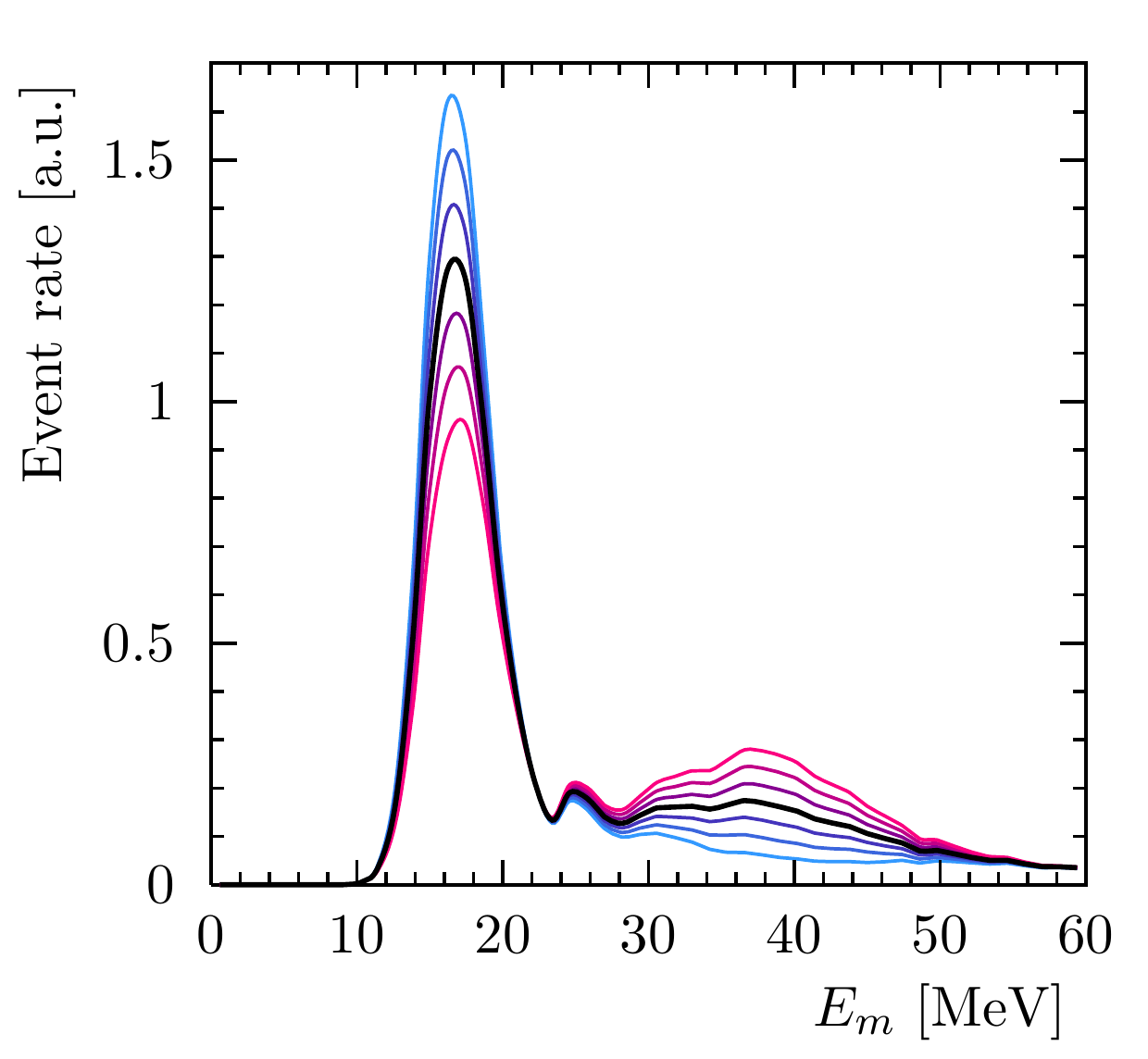}
    \includegraphics[width=0.95\linewidth]{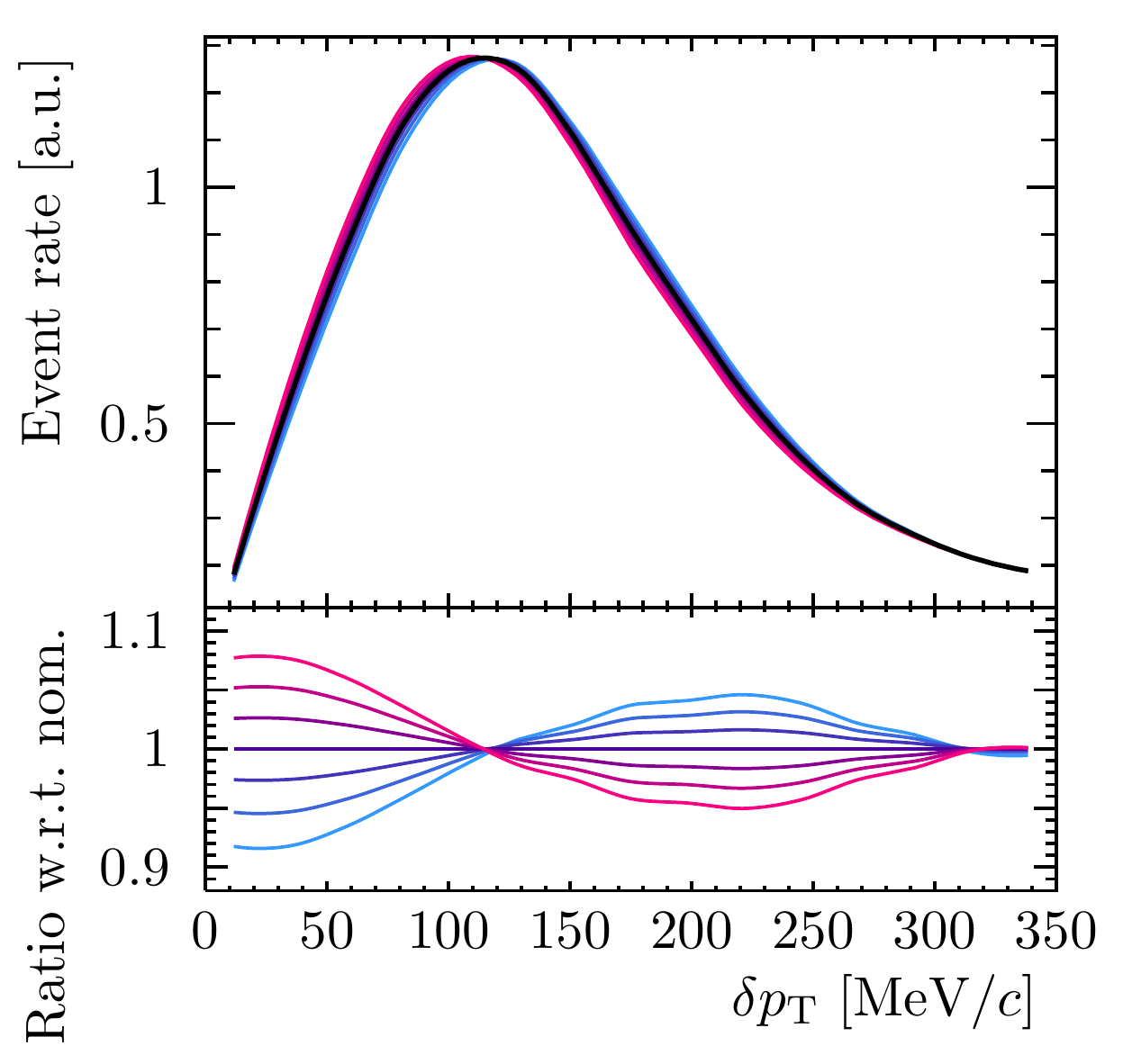}
    \caption{The distributions for removal energy (top) and missing transverse momentum (bottom) as given by the SF model implemented in \texttt{NEUT}, showing the impact of the shell normalization parameters ($N_\mathrm{shell}$) from $-1.5\sigma$ (pink) to $+1.5\sigma$ (cyan), compared to the nominal (black). The interactions were generated on a carbon target using the T2K flux.} 
    \label{fig:shellnorm}
\end{figure}

As shown in Fig.~\ref{fig:2DSFNEUT}, the missing energy distributions in the SF model exhibit multiple peaks, which correspond to the nucleon energy levels in a shell model. Electron scattering measurements show that the relative strength of each shell may differ from those predicted by a simple independent-particle shell model~\cite{Gonzalez-Jimenez:2021ohu, Franco-Patino:2022tvv}. To account for this, a parameter that modifies the shell occupancy is introduced as a normalization uncertainty for each shell in the MF region of the SF model. 
Weights are assigned to CCQE events in this region following:
\begin{eqnarray}
    f_{\mathrm{shell}} (E_m) &=& 1 + N_{\mathrm{shell}} \times \exp \left( - \frac{(E_m - E_{\mathrm{shell}})^2}{2 \sigma_{\mathrm{shell}}^2} \right)\\
    &=& 1 + N_{\mathrm{shell}} \times g_{\mathrm{shell}}(E_m)
\end{eqnarray}
where $N_{\mathrm{shell}}$ is the normalization parameter of a given shell, and $E_{\mathrm{shell}}$ and $\sigma_{\mathrm{shell}}$ correspond to the center and the width of the Gaussian function, $ g_{\mathrm{shell}}(E_m)$, which are fixed for each shell. In total, this gives two shell normalization parameters for interactions with carbon and three for interactions with oxygen. The fixed values of $E_{\mathrm{shell}}$ and $\sigma_{\mathrm{shell}}$ are derived from an analysis of the missing energy distributions in \texttt{NEUT}, and are shown for each shell in Tab.~\ref{tab:num}. The impact of varying $N_{\mathrm{shell}}$ for the carbon $p$- and $s$-shells is shown in Fig.~\ref{fig:shellnorm}.

As can be seen in the $E_m$ distribution from Fig.~\ref{fig:shellnorm}, this approach results in a thinner or thicker shell for higher or lower $N_{\mathrm{shell}}$ values respectively and allows the normalization of the cross section to change (i.e.\ the total spectral function normalization is not scaled down to account for an increase in a particular $N_{\mathrm{shell}}$). Note also that there is no limit to the range of $E_m$ each shell normalization uncertainty dial is applied to. Each event gets weights from shell occupancy dials for every shell, but if an event is far away from a shell center then the corresponding shell occupancy weight would require very large values of $N_{\mathrm{shell}}$ to give the event a significant weighting.

\bgroup
\def\arraystretch{1.3}
\begin{table}[ht]
    \centering
    \begin{tabular}{|c|c|c|c|c|}
        \hline
        Target & Shell & $E_{\mathrm{shell}}$ [MeV] & $\sigma_{\mathrm{shell}}$ [MeV] & $N_{\mathrm{shell}}$ uncertainty \\
        \hline 
        \multirow{2}{*}{Carbon} & $p$ & 18 & 15 & 0.2\\ 
         & $s$ & 36 & 25 & 0.4\\
        \hline 
        \multirow{3}{*}{Oxygen} & $p_{1/2}$ & 12 & 8 & 0.45\\ 
         & $p_{3/2}$ & 19 & 8 & 0.25\\
         & $s$ & 42 & 25 & 0.75\\
        \hline
    \end{tabular}
    \caption{Nuclear energy levels with their widths for the different shells, for the SF as implemented in \texttt{NEUT}. The last column represents the relative prior uncertainty set on the corresponding shell normalization parameter, which all have a central value of 0.}
    \label{tab:num}
\end{table}
\egroup

\begin{figure}[]
    \centering
        \includegraphics[width=0.95\linewidth]{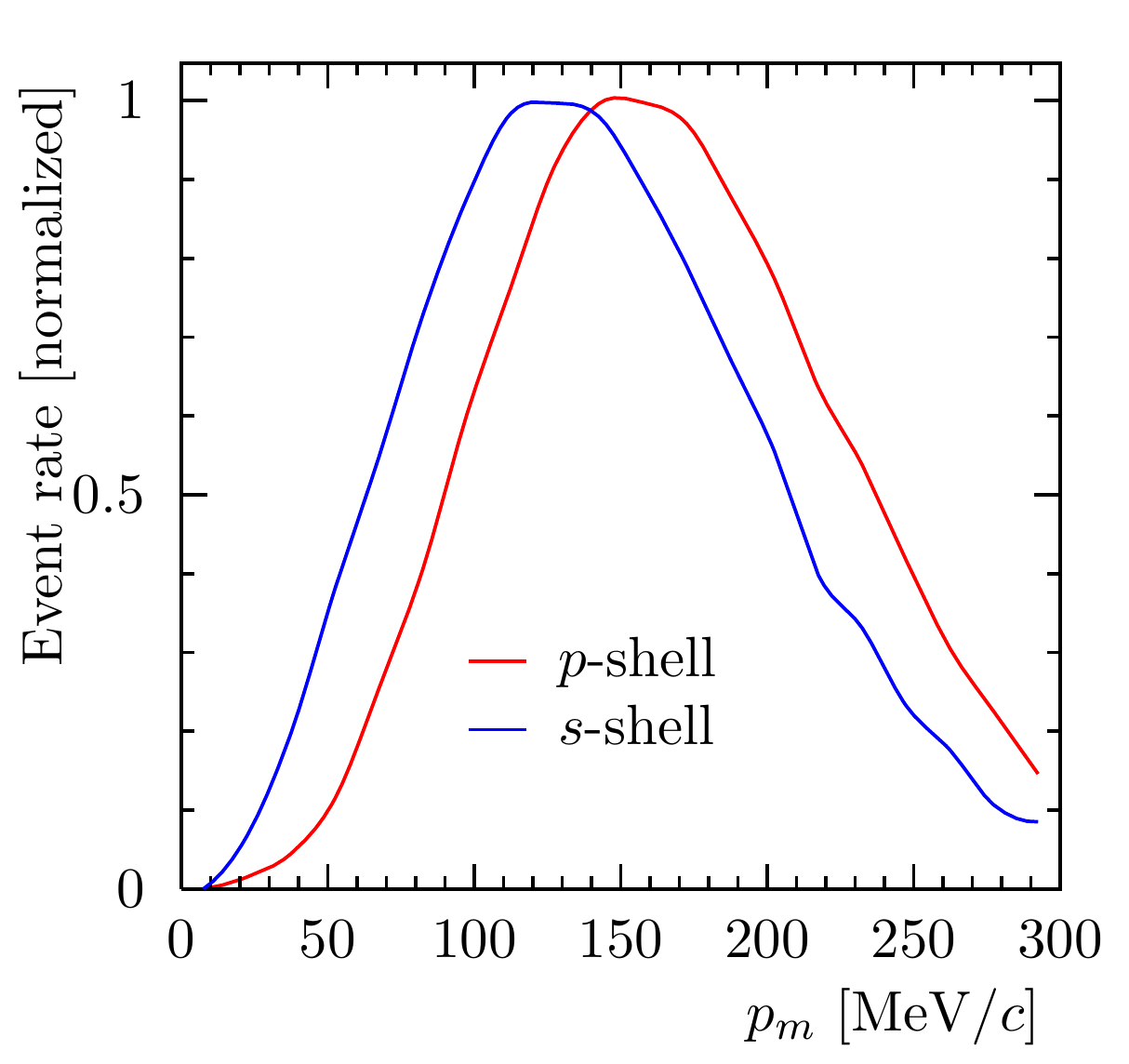}
    \caption{
    Distribution of the missing momentum within each nuclear shell for carbon as given by the SF model implemented in \texttt{NEUT} generated using the T2K flux.}
    \label{fig:shelldist}
\end{figure}

One important effect of the shell occupancy parameters is how they alter the total $p_m$ distribution, since the initial-state nucleon's momentum distribution differs between the shells, as shown in Fig.~\ref{fig:shelldist}. Therefore, a change in the relative strength of a shell impacts the shape of the overall distribution of $p_m$, and consequently the distribution of \dpt, as is illustrated in the lower panel of Fig.~\ref{fig:shellnorm}. Such freedom allows for variation of the outgoing lepton and nucleon kinematics in a way that is consistently propagated through the model.

\begin{figure}[]
    \centering
    \includegraphics[width=0.95\linewidth]{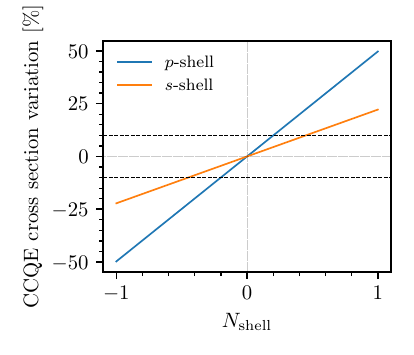}
    \includegraphics[width=0.95\linewidth]{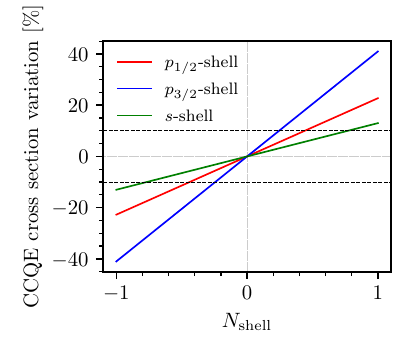}

    \caption{The impact of the shell normalization parameters ($N_\mathrm{shell}$) on the total CCQE cross section when applied to the SF model implemented in \texttt{NEUT} for each shell in carbon (top) and oxygen (bottom). The dashed horizontal lines indicate the $\pm 10\%$ variations, chosen to correspond to the $1\sigma$ uncertainty for these parameters. The interactions were generated using the T2K flux.}
    \label{fig:shellnorm_vs_xsec}
\end{figure}

As discussed, we opt to keep the prior uncertainties on the $N_{\mathrm{shell}}$ parameters conservative. The chosen values are reported in Tab.~\ref{tab:num}. These values cover differences in the shell occupancy of the SF model (derived from $(e,e'p)$ data) to those expected in an independent particle shell model~\cite{Gonzalez-Jimenez:2021ohu, Franco-Patino:2022tvv}, whilst ensuring the same variation for each parameter changes the total CCQE cross section by the same amount ($\sim10\%$). 
Fig.~\ref{fig:shellnorm_vs_xsec} shows the impact of varying each shell normalization parameter $N_{\mathrm{shell}}$ on the total CCQE cross section, and Fig.~\ref{fig:shellnorm} demonstrates potential sensitivity to variations through measurements of \dpt shape. If the parameters are pulled far outside the prior uncertainties, this could indicate that the parameters are acting as effective degrees of freedom to account for physics beyond the PWIA. 

\subsubsection{Missing-momentum shape}

\begin{figure*}[]
    \centering
    \includegraphics[width=0.45\linewidth]{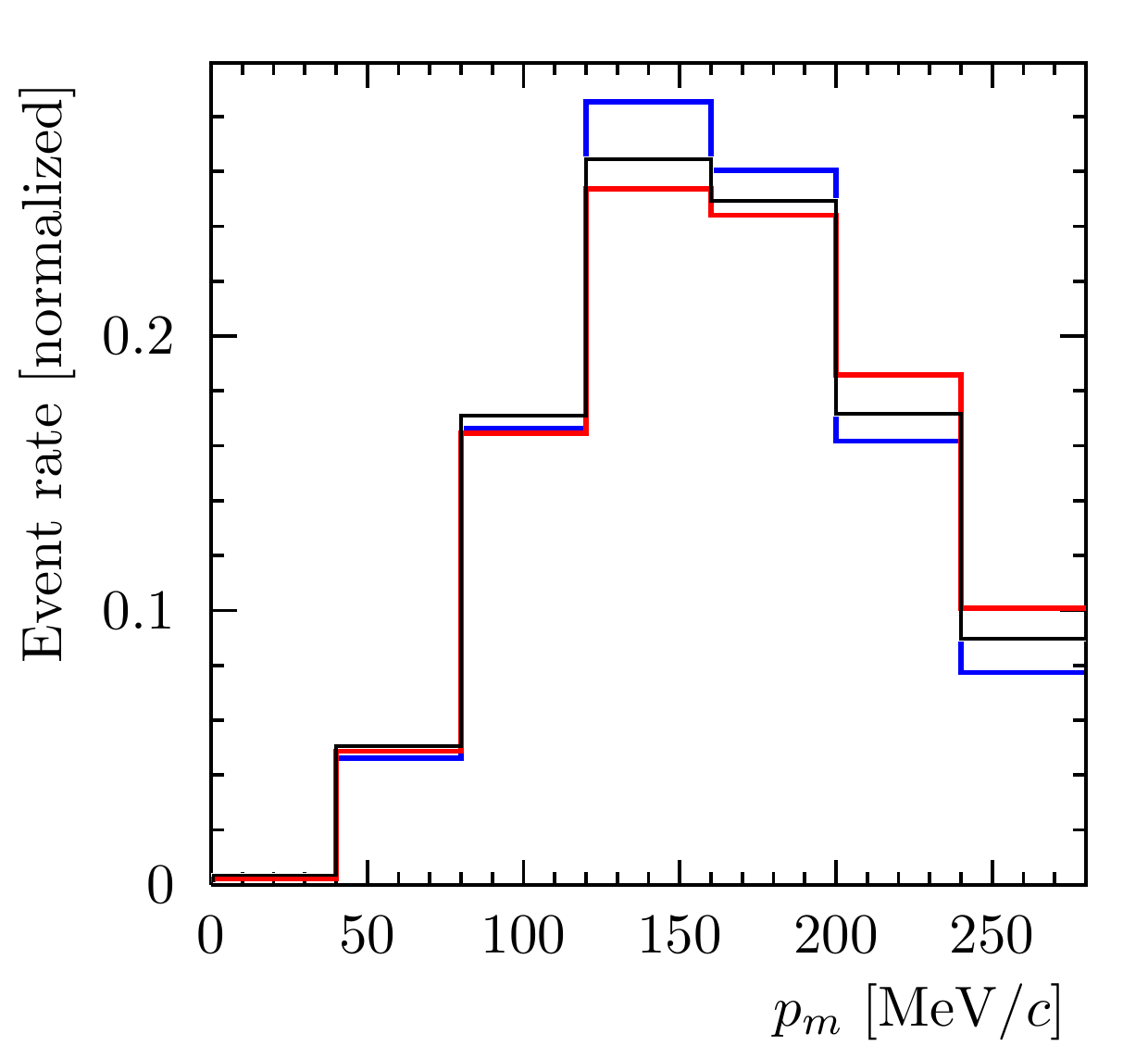}
    \includegraphics[width=0.45\linewidth]{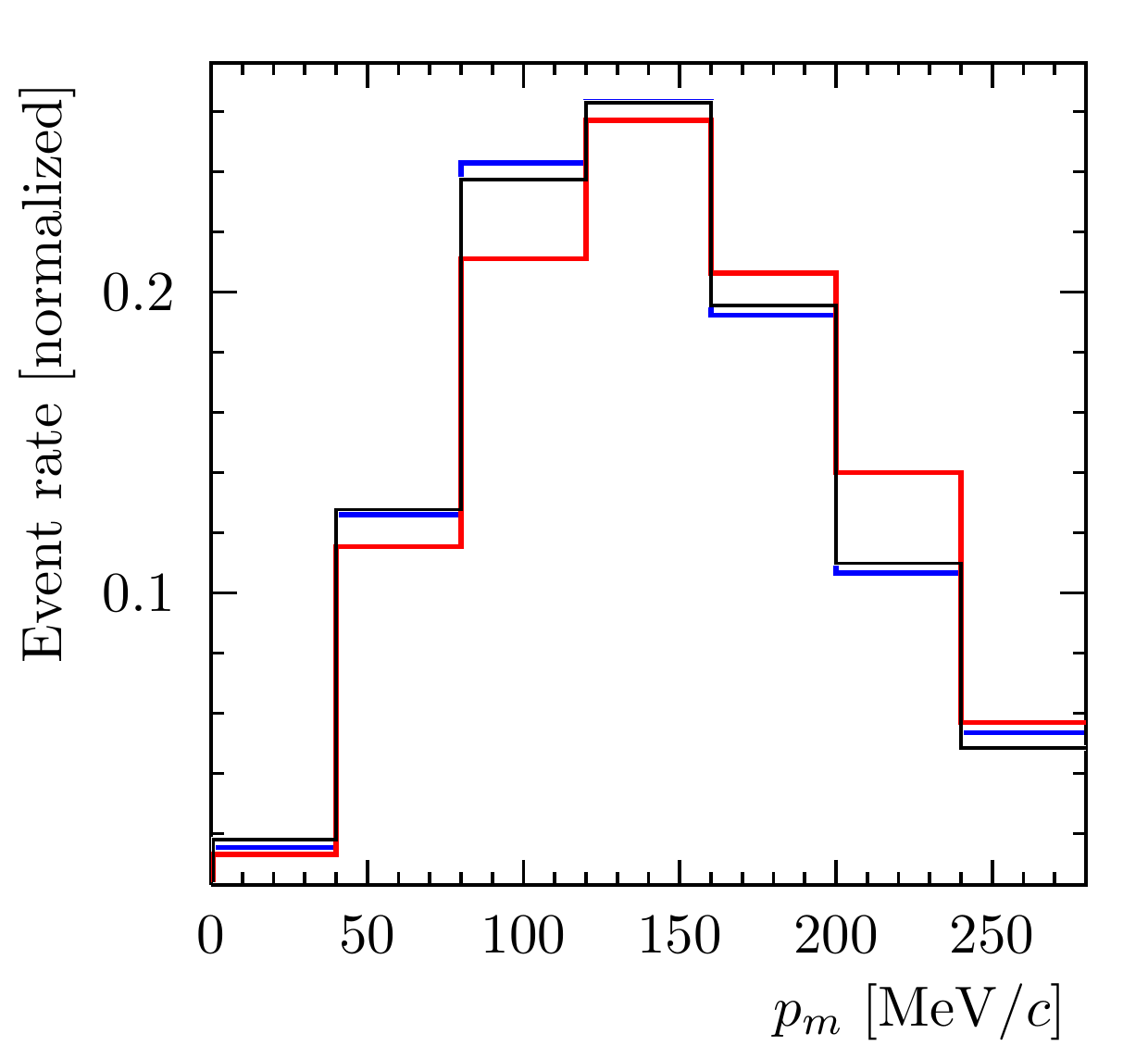}
    \caption{Distributions of the missing momentum from the \texttt{NEUT} SF inputs (black) compared to electron scattering measurements~\cite{JLabE91013:2003gdp} (blue and red), made for different nucleon and lepton kinematics for carbon in the $p$-shell (left) and the $s$-shell (right).}
    \label{fig:neutpmissvsdata}
\end{figure*}

Electron scattering data from Ref.~\cite{JLabE91013:2003gdp} provides measurements of the missing momentum distributions $p_m$ for carbon. By comparing the predicted \texttt{NEUT} distributions of $p_m$ in each shell (which are taken directly from Benhar SF model predictions) with this data, small shape differences appear, as illustrated in Fig.~\ref{fig:neutpmissvsdata}. This shows two ``extreme'' distributions of the measured missing momentum which correspond to $(e,e'p)$ kinematics with the most different four-momentum transfer, $Q^2$, compared to the \texttt{NEUT} prediction. This builds a missing-momentum shape uncertainty for each shell that changes the shape of the $p_m$ distribution. 
Each shell parameter is defined between -1 and 1, corresponding to the two extreme $p_{m}$ distributions, and a linear extrapolation is implemented beyond the range $[-1, 1]$. 

\begin{figure*}[]
    \centering
    \includegraphics[width=0.45\linewidth]{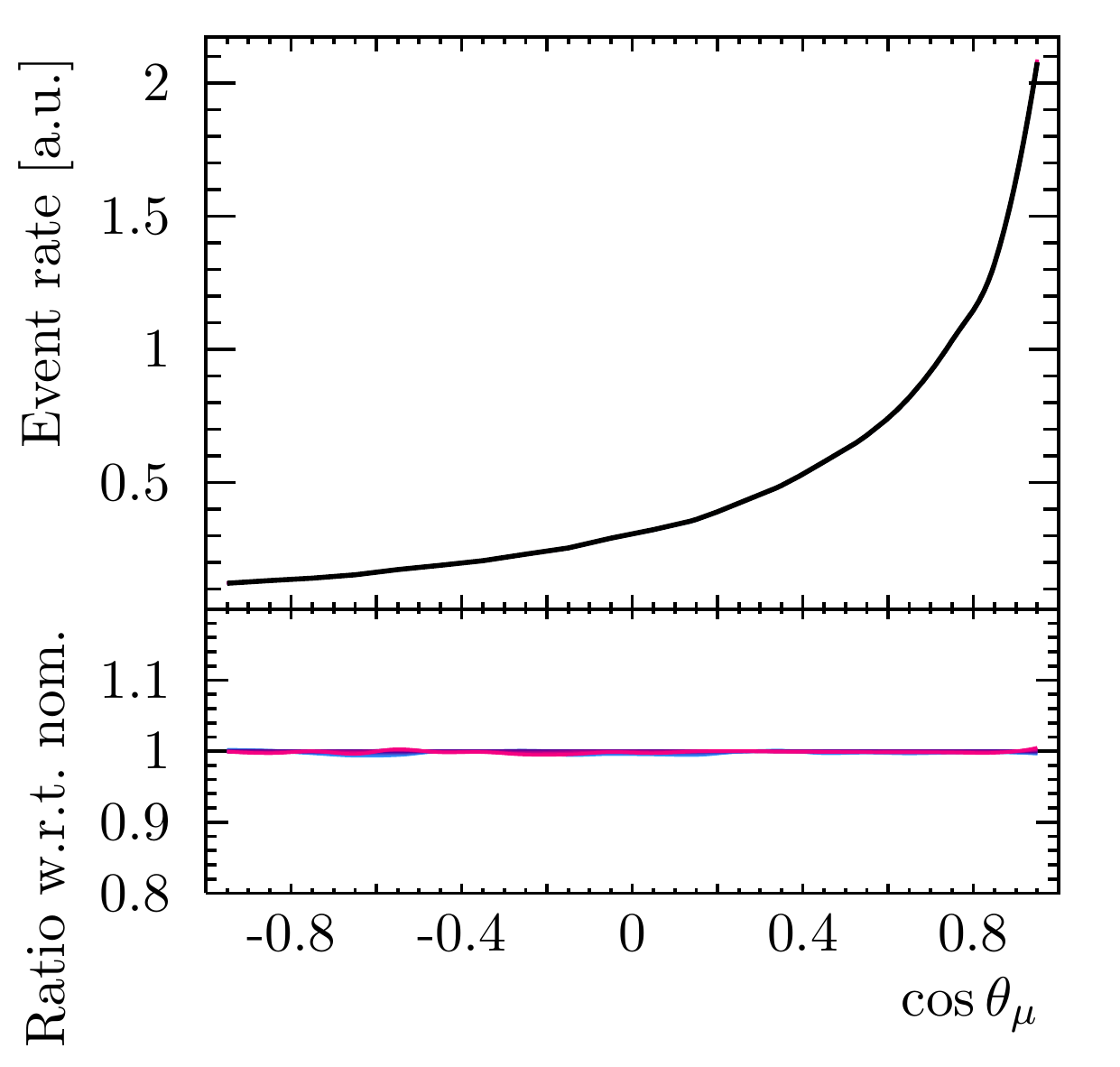}
    \includegraphics[width=0.45\linewidth]{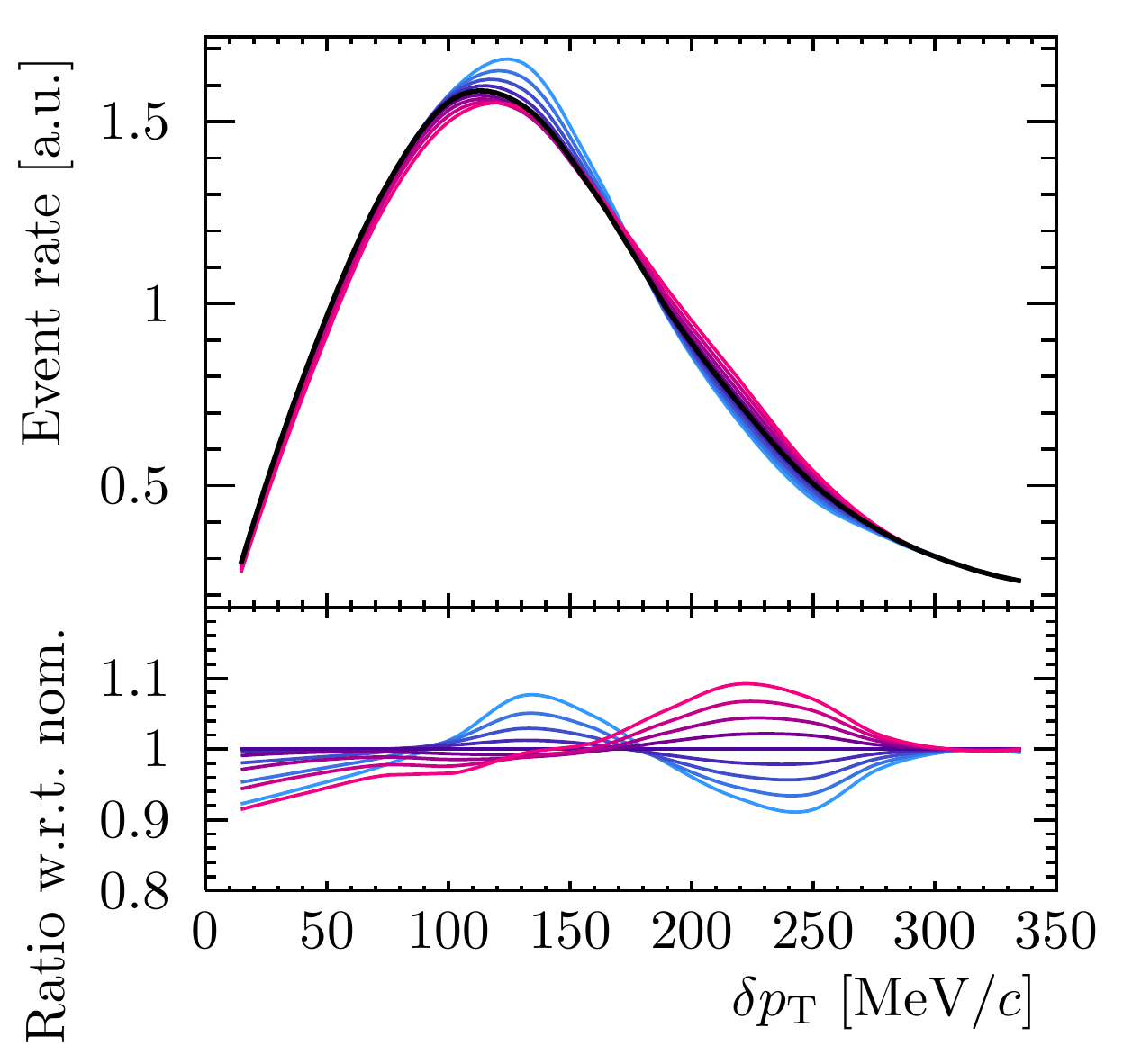}
    \caption{The distributions for the angle between the outgoing lepton and neutrino (left) and missing transverse momentum (right) as given by the SF model implemented in \texttt{NEUT}, showing the impact of variations of the $p$-shell shape uncertainty from $-2\sigma$ (pink) to $2\sigma$ (cyan), compared to the nominal (black). The interactions were generated on a carbon target using the T2K flux.}
    \label{fig:pshell_dpt_cthmu}
\end{figure*}

These $p_m$ shape parameters are expected to mainly affect experimental observables sensitive to the initial-state nuclear momentum. This is demonstrated in Fig.~\ref{fig:pshell_dpt_cthmu}, which shows the impact of the $p$-shell shape parameter on the distributions of the neutrino-muon angle and the transverse momentum imbalance. The lepton kinematics have no discernible sensitivity to these variations, whereas the bulk of \dpt is affected by this uncertainty since it corresponds, in the absence of other nuclear effects, to the transverse projection of the Fermi motion. 

\subsubsection{Short-range correlations}

As previously discussed in Sec.~\ref{sec:SF}, the SRC contribution implemented in \texttt{NEUT} and \texttt{NuWro} neutrino event generators corresponds to CCQE events yielding two outgoing nucleons following the primary interaction. 
Calculations performed by Benhar \etal~\cite{Benhar:1994hw} are implemented in neutrino event generators via tables of the total SF distribution containing both the MF and SRC components. Whilst these are separate components of the original SF calculation, the information on their shape is not preserved in the generator implementations. Therefore, it is necessary that generators develop a scheme to separate the two components. 
\texttt{NEUT} uses hard cuts on $E_m$ and $p_m$ to distinguish between them as shown in Fig.~\ref{fig:2DSFNEUT}: an SRC two-nucleon knock-out only occurs if the neutrino interacts with a nucleon for which $E_m > 100$~MeV or $p_m > 300$~MeV/$c$. The spectator nucleon of the pair is taken to have opposite isospin and momentum compared to the ``active'' interacting nucleon. 
Using this implementation, \texttt{NEUT} predicts that interactions with SRC pairs represent $\sim 5\%$ of the total CCQE interactions for both carbon and oxygen. 
Alternatively, the SF implementation in \texttt{NuWro} takes a different approach by making non-rectangular cuts in the $(p_m, E_m)$ phase space adapted to each target in a more phenomenological manner, also shown in Fig.~\ref{fig:2DSFNEUT}. 
Furthermore, while the hard cuts in \texttt{NEUT} fully determine the MF-SRC separation, \texttt{NuWro} applies an additional condition to allow for the knock-out of the SRC pair. It requires the energy of the pair to be higher than 14~MeV, i.e.\ approximately twice the average nucleon removal energy. As a consequence, the SF model implementation in \texttt{NuWro} predicts a larger SRC contribution, amounting to $\sim$15\% of the total CCQE interactions. 

\begin{figure*}[]
    \centering
        \includegraphics[width=0.45\linewidth]{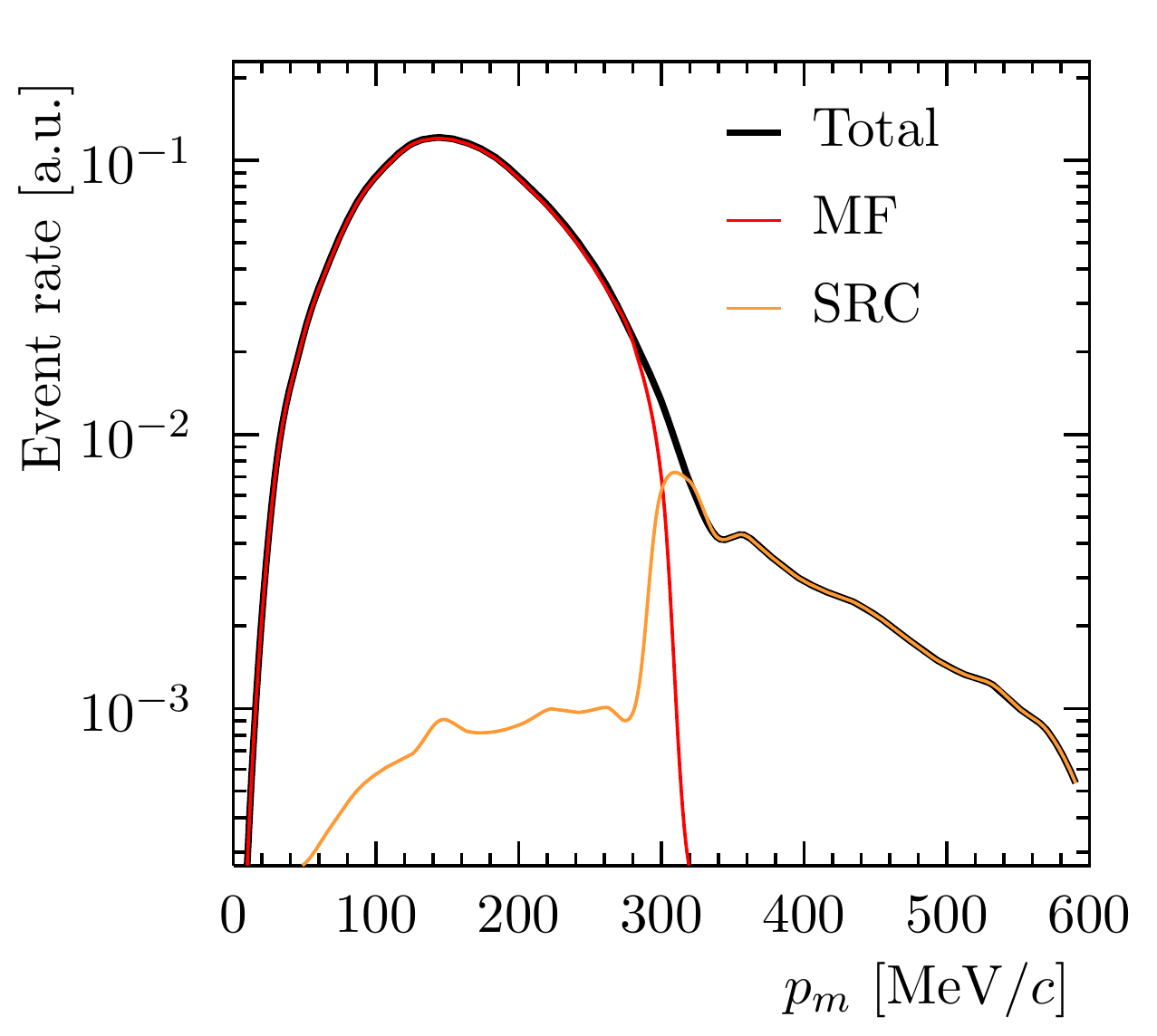}
        \includegraphics[width=0.45\linewidth]{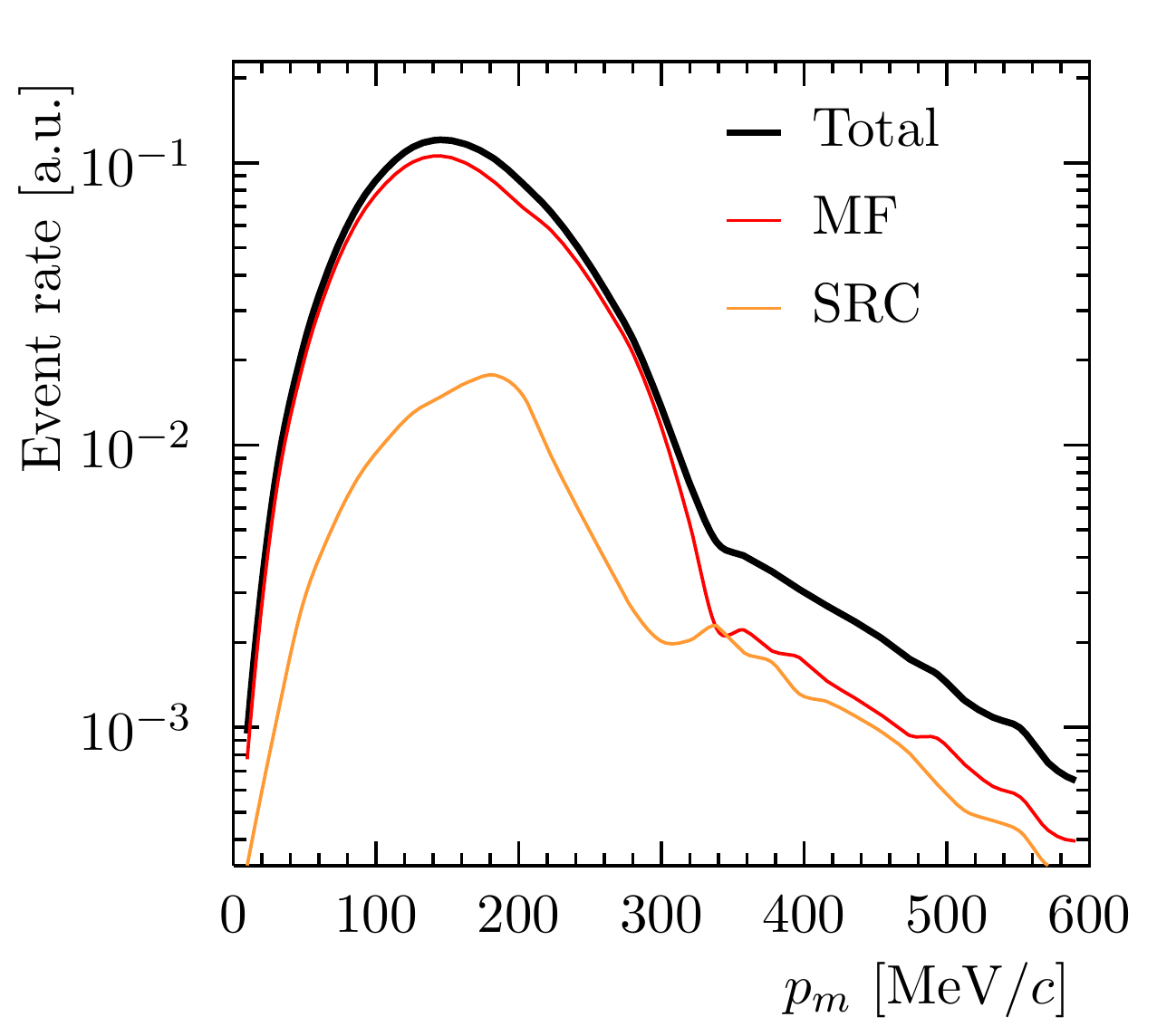}
    \caption{Distribution of the missing momentum in the SF model in \texttt{NEUT} (left) and \texttt{NuWro} (right) for carbon with the MF and SRC contributions.}
    \label{fig:pmissSRC}
\end{figure*}

The impact of these different implementation choices in \texttt{NEUT} and \texttt{NuWro} on the missing-momentum distribution is displayed in Fig.~\ref{fig:pmissSRC}. 
While the high $p_m$ tail in \texttt{NEUT} is exclusively due to SRC, it is not necessarily the case in \texttt{NuWro} because of the additional condition on the energy of the SRC pair. 
These clear differences motivate the need for a large uncertainty on the SRC contribution to the cross section.
This is applied as a normalization parameter of SRC events. We set its prior uncertainty to a value of 100\%. Although it should be noted that this uncertainty does not cover the full $p_m$ shape differences in Fig.~\ref{fig:pmissSRC}. Improved SRC uncertainties would require a more theory-driven SRC implementation which contributes significantly across the full range of $p_m$ (to allow phase space coverage for event reweighting). It should also be noted that the SRC alteration parameter allows modification of the total CCQE cross section. There is no attempt in this analysis to offset alterations to the SRCs by modifications to the MF-region shell normalizations.

\subsection{Uncertainties for physics beyond PWIA}
\label{sec:pwiauncert}

In addition to parameters to alter the natural degrees of freedom within the SF model, additional freedoms are required to account for plausible variations from the PWIA the model is built on. The \texttt{NEUT} implementation of PB, as described in Eq.~\ref{eq:PB}, provides a simple freedom to vary its treatment, whilst additional uncertainties on the alteration to the cross-section at low energy transfers due to FSI effects can also be considered.

\subsubsection{Pauli blocking}
As discussed in Sec.~\ref{sec:SF}, the SF model in \texttt{NEUT} features a simple description of Pauli blocking (PB) inspired by the RFG model of the nuclear ground state~\cite{Hayato:2021heg}. With this approach, events with an outgoing primary nucleon with momentum below the Fermi surface momentum, $p_F$, are removed and do not contribute to the total cross section. In \texttt{NEUT}, $p_F$ is set to 209 MeV/$c$ for both carbon and oxygen. PB both reduces the cross section predicted by the SF model and causes significant shape changes for events with low momentum outgoing nucleons, often corresponding to low momentum transfer.
Analyses of $p_F$ from theory and electron scattering data span a range of $\sim$20 MeV/$c$~\cite{Benhar:2005dj,Ankowski:2014yfa,PhysRevLett.26.445, PhysRevC.65.025502}, although we opt for a more conservative size of uncertainty, given the simplistic approach of the PB discussed earlier. Consequently, a parameter varying the threshold $p_F$ is prescribed separately for each nuclear target with a conservative prior uncertainty of $\pm30$~MeV/$c$. Note that \texttt{NEUT} simulates Pauli blocking effects both as part of the primary neutrino interaction and then again as part of its intranuclear FSI cascade simulation. The parameter introduced here only affects the former. Uncertainties related to variations of the FSI cascade are discussed in Sec.~\ref{sec:FSI}.


\begin{figure}[]
    \centering
    \includegraphics[width=0.95\linewidth]{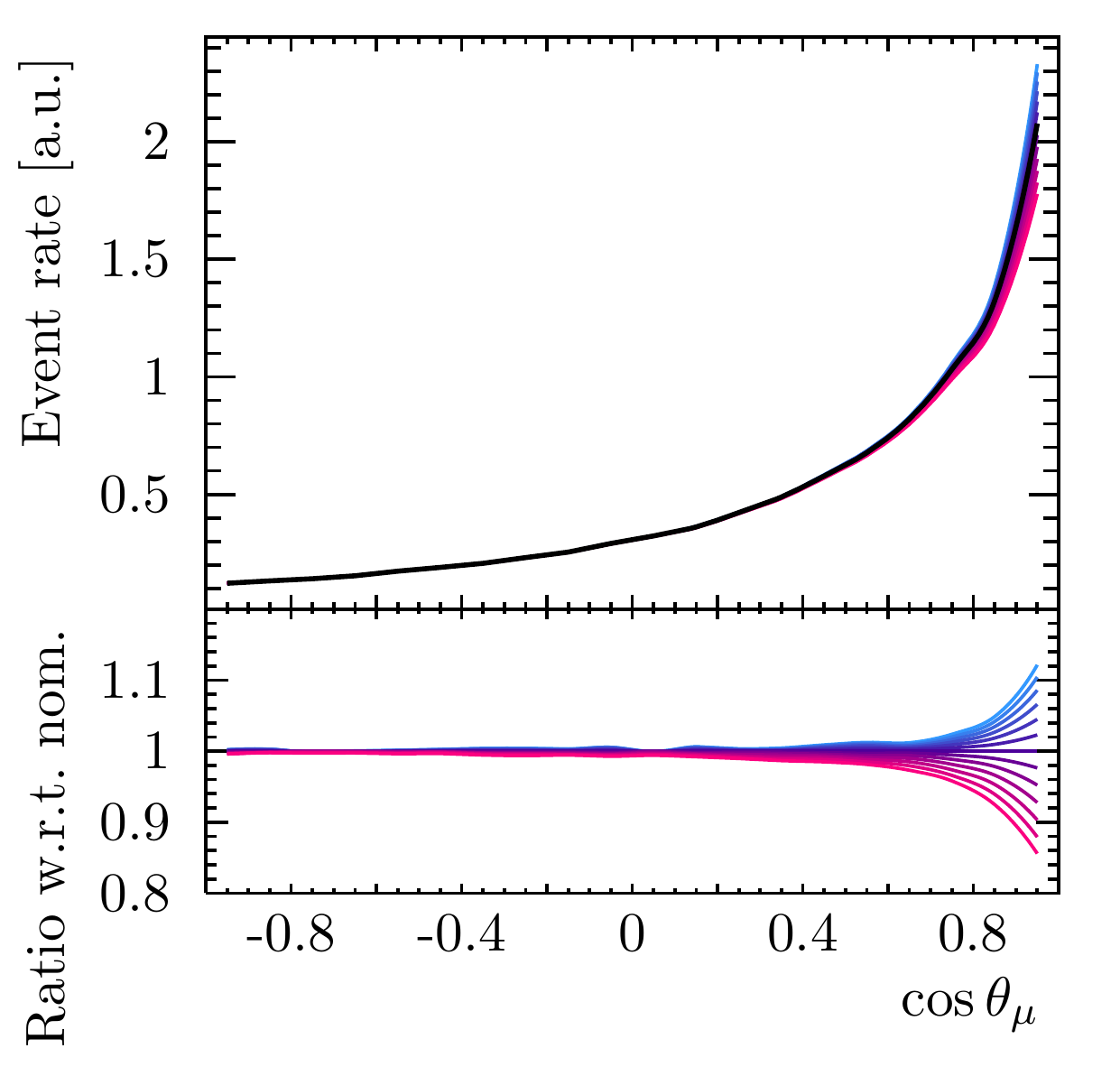}
    \includegraphics[width=0.95\linewidth]{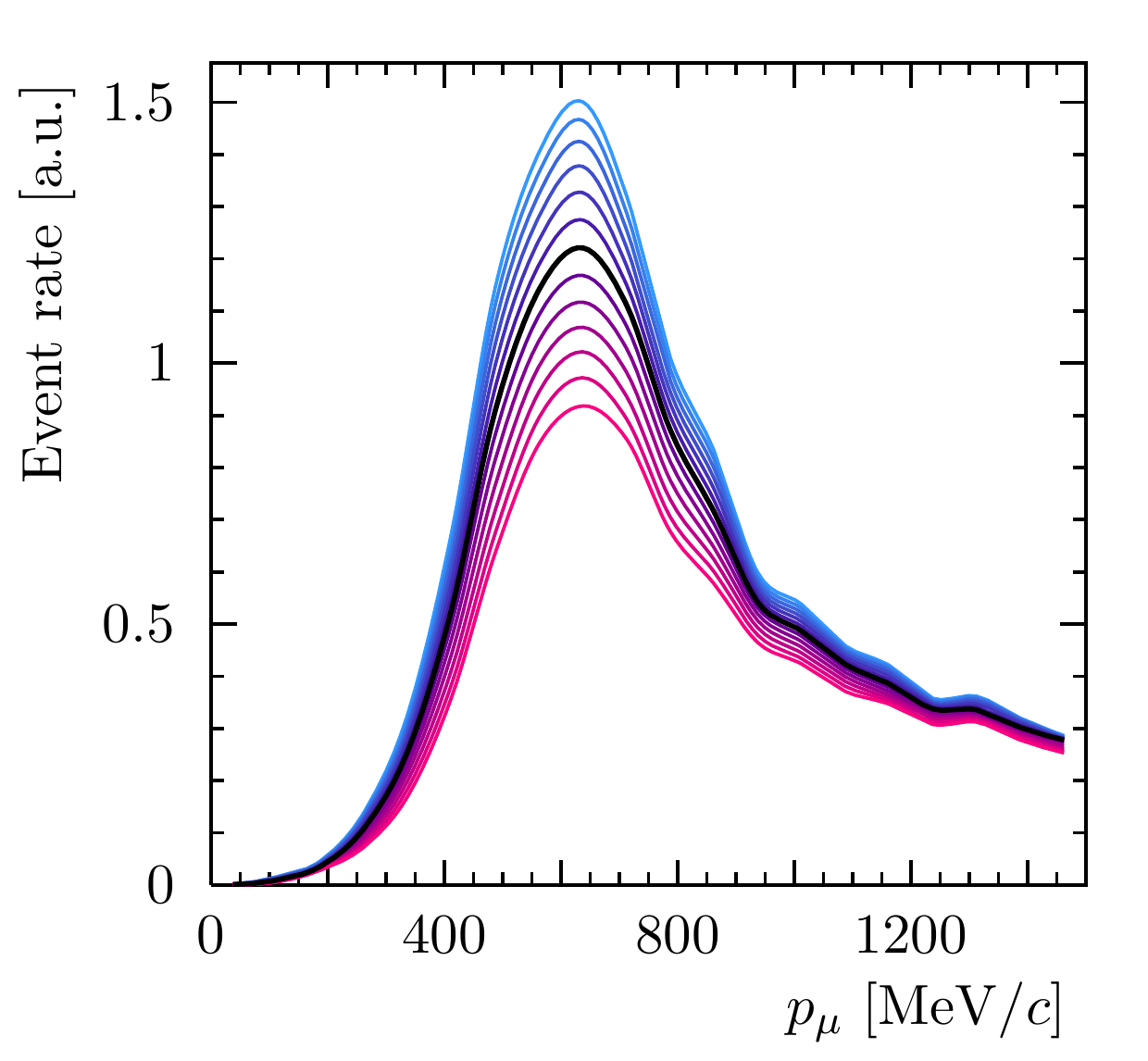}
    \caption{The distributions of the angle between the outgoing muon and neutrino (top) and the muon momentum in the $\cos \theta_\mu > 0.9$ region (bottom), showing the impact of varying the PB threshold from $-1.5\sigma$ (cyan) to $+1.5\sigma$ (pink), compared to the nominal (black). The interactions were generated on a carbon target using the T2K flux.}
    \label{fig:PB}
\end{figure}

Fig.~\ref{fig:PB} shows the impact of varying the PB threshold parameter on the lepton kinematics. Since PB only impacts events with outgoing nucleons with low momentum, its impact is most prominent at low energy transfers. Therefore, PB is most noticeable for forward-going leptons. The impact of varying PB is not significant in semi-inclusive measurements because low momentum nucleons are seldom detected due to the high thresholds in current detectors ($p_{p}\sim450~\textrm{MeV}/c$).

\subsubsection{Optical potential}
The addition of an intranuclear cascade to simulate FSI distorts the outgoing nucleon momentum distribution and accounts for additional hadron ejection, but does not change the inclusive CCQE cross section. A full treatment of the distortion of the outgoing nucleon wave-function would affect the cross section as a function of both lepton and nucleon kinematics~\cite{Nikolakopoulos:2022qkq}. To account for this missing alteration to the inclusive cross section due to FSI--which is physics beyond the PWIA--Ref.~\cite{Ankowski:2014yfa} calculates a correction based on an optical potential (OP) that is derived as a correction to the SF model's prediction of the lepton kinematics. 

Although such effects are not implemented in \texttt{NEUT}, \texttt{NuWro v19.02.01} has an option to include this correction. \texttt{NuWro} was thus used to define an OP parameter that varies the cross section in energy and momentum transfer $(q_0, \left| \vec q\right|)$. The parameter is the amount of the OP effect on the nominal \texttt{NEUT} MC, from $0\%$ (no correction) to $100\%$ (full correction), using a linear interpolation between the two. The calculation is only available for carbon, although it is not expected to be dramatically different for oxygen, so the parameter is applied independently for each target. To avoid fit convergence issues at the boundaries of this parameter, the prior central value is set to $50 \%$ with a $\pm 50\%$ uncertainty. Hence, the prior acts to stabilize the fit whilst being conservative enough that any constraint in the fit is dominated by those implied by cross-section measurement. 

\begin{figure}[]
    \centering
    \includegraphics[width=0.95\linewidth]{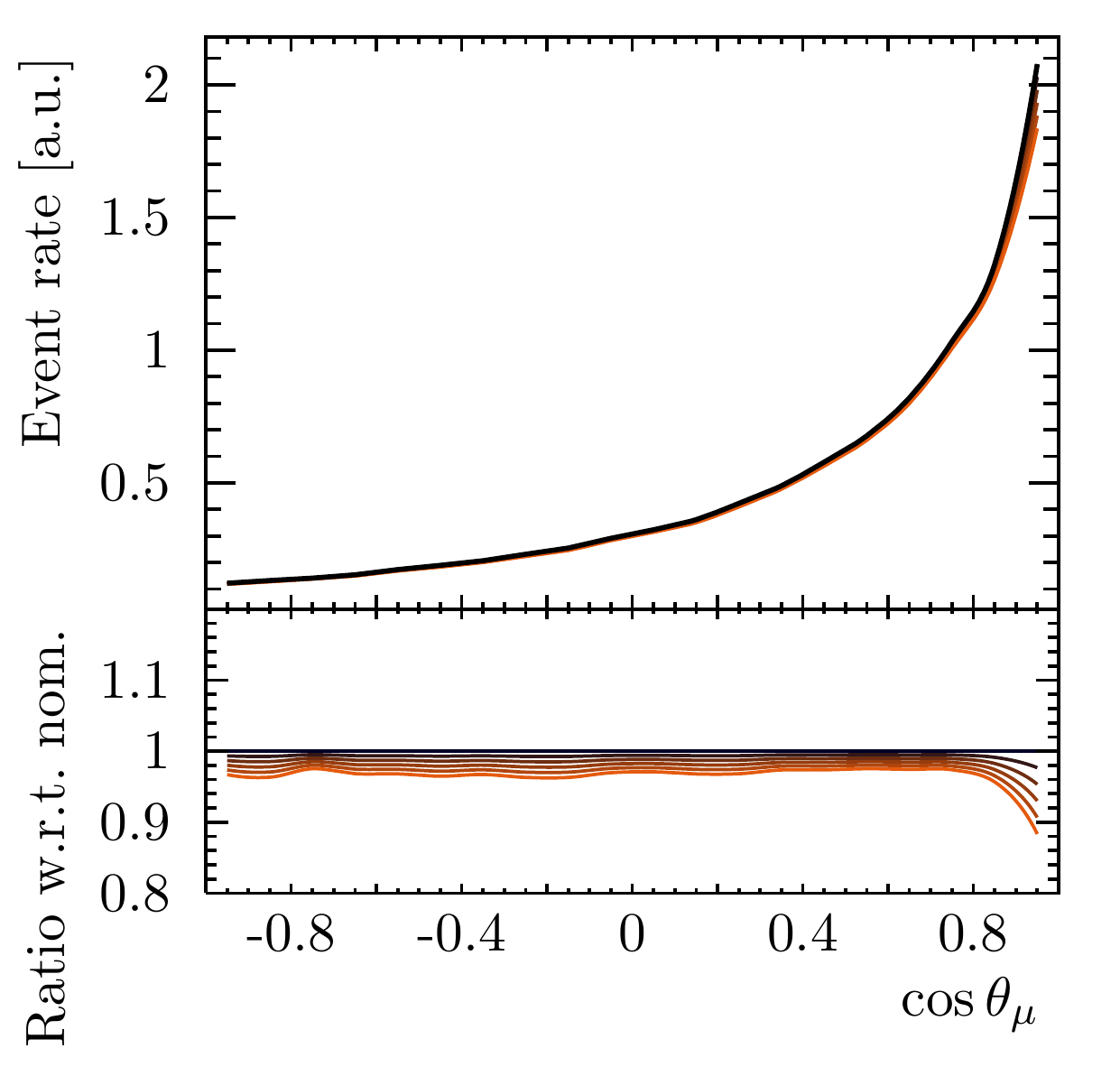}
    \includegraphics[width=0.95\linewidth]{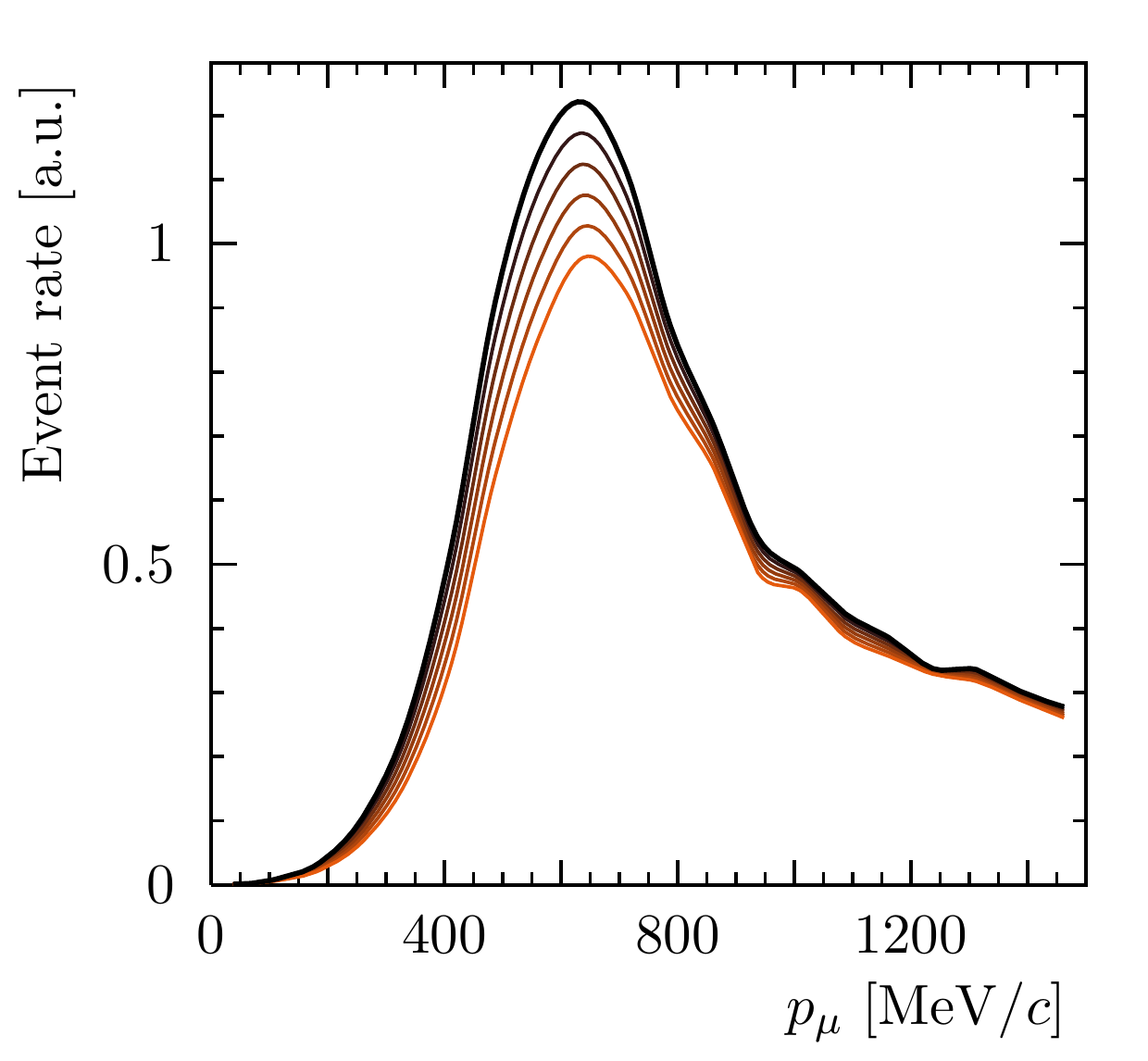}
    \caption{The distributions of the angle between the outgoing muon and neutrino (top) and the muon momentum in the $\cos \theta_\mu > 0.9$ region (bottom), showing the impact of varying the PB threshold from from $0\%$, i.e.\ nominal, (black) to $100\%$ (orange), compared to the nominal (black).}
    \label{fig:OP}
\end{figure}

As discussed in Sec.~\ref{sec:SF}, effects beyond the PWIA appear at low energy and momentum transfer, where the bound nucleons can no longer be treated as independent entities. Similarly to PB, Fig.~\ref{fig:OP} illustrates the impact of applying the OP correction on the lepton kinematics. The largest impact is again for forward-going leptons, since the transferred momentum and energy in such interactions is the smallest. Whilst the single-dimensional projections show the impact of OP and PB parameters to be similar, they are not degenerate in higher dimensions: for example, their impact on muon momentum for fixed ranges of muon angle are different.


Since the impact of this correction is calculated only for the outgoing lepton kinematics, the method of applying it to the \texttt{NEUT} distribution of $(q_0, \left| \vec q \right|)$ from \texttt{NuWro} only alters kinematics of the outgoing nucleon via the lepton-nucleon correlations in the implemented SF model. A full treatment of an OP correction would alter these correlations, and so it is expected that the SF model with the OP correction may not have strong predictive power for outgoing nucleon kinematics. An additional uncertainty on the nucleon kinematics is implemented via uncertainties in the FSI cascade is applied, which allows some freedom to mitigate this issue and is discussed in Sec.~\ref{sec:FSI}.

\subsection{Additional CC0$\pi$ uncertainties}\label{sec:other}
To benchmark and validate the uncertainty parametrization of the SF model described in the previous sections, cross-section measurements from both the T2K and \minerva collaborations are fit, as is discussed in Sec.~\ref{sec:fit}. The topology of interest is the charged-current interactions without pions in the final state, known as CC0$\pi$, since it provides a sample enriched with CCQE events, thus relevant for studying the SF model. This topology also contains significant fractions of 2p2h and single-pion production events with a subsequent absorption of the outgoing pion, which thereby necessitate additional systematic uncertainties to compare to the measurements. The non-CCQE contributions relevant to the T2K and \minerva measurements are different due to their different neutrino energy range. T2K has a narrow-band neutrino energy spectrum with a peak neutrino energy of $E_\nu\sim0.6~\textrm{GeV}$, and \minerva has a wide-band beam with average neutrino energy $\langle E_\nu \rangle \sim3.5~\textrm{GeV}$.


\subsubsection{CCQE axial form factor}
Since the treatment of the axial form factor can significantly impact the CCQE cross section---especially at high four-momentum transfers---we also consider the systematic uncertainty related to the dipole form of the axial form factor, typically used in neutrino event generators. The value of the nucleon axial mass $M_A^\mathrm{QE}$ in the dipole form factor is set to $1.03\pm0.06~\textrm{GeV}/c^2$, estimated from analysis of bubble chamber data~\cite{T2K:2023smv}. 
Additionally, the dipole approximation of the form factor may be insufficient to describe the evolution of the cross section, particularly at high $Q^2$~\cite{Meyer:2022mix}. 
Therefore, three ``high $Q^2$ parameters'', which vary the normalization of CCQE interactions in the ranges $Q^2=[0.25-0.50], [0.50-1.00]$, and $[1.00-\infty]$ in GeV$^2$, are applied to allow freedom beyond the dipole form factors. This approach is derived from the $Q^2>0.25$~GeV$^2$ parameters in Ref.~\cite{T2K:2023smv}. As also detailed in Ref.~\cite{T2K:2023smv}, the corresponding uncertainties are derived from comparisons of the shape in the high $Q^2$ region of the dipole and other models, including the z-expansion model~\cite{Bhattacharya:2011ah,Meyer:2016oeg}.

\subsubsection{Nucleon final-state interactions}
\label{sec:FSI}
As discussed, FSI play a crucial role in altering the outgoing nucleon kinematics and distorting interaction topologies. This is especially relevant when the cross-section measurements are performed in variables that use both the lepton and the hadron information, such as \dpt. To account for nucleon FSI uncertainty in this work, CC$0\pi$ events are divided into two classes: 
\begin{itemize}
    \item ``{With FSI}'': when the outgoing nucleon kinematics are modified due to FSI,
    \item ``{Without FSI}'': in the opposite case, when the nucleon exists the nucleus without being impacted by the intranuclear cascade.
\end{itemize}
A normalization parameter for each class is applied with a broad $30\%$ prior uncertainty, motivated by an analysis of nuclear transparency measurements~\cite{Niewczas:2019fro}. 
The two parameters are very strongly anti-correlated to ensure that the total cross section remains almost constant for each interaction mode. If the amount of ``{Without FSI}'' events in a certain interaction mode is reduced, the amount of ``{With FSI}'' events in that same interaction mode is therefore increased, yielding a shape-only effect. Effectively these two parameters act almost as a single degree of freedom and are only split for technical implementation reasons,
which has a similar impact as changing the mean free path of the nucleon within the nucleus. 
It should be noted that this approach does inadvertently allow some small changes in shape to the cross section as a function of lepton kinematics, which should remain unchanged by FSI cascades.
This is an extremely simple parametrization of FSI uncertainties which is unlikely to be sufficient for a detailed analysis of outgoing hadron kinematics, but can be suitable for analyses that are only sensitive to the coarse hadron kinematic information considered in this work. Improvement of FSI cascade uncertainty parametrizations is a crucially important topic for modeling neutrino interactions in active development (see e.g.\ Refs~\cite{Nikolakopoulos:2022qkq,Ershova:2022jah,Isaacson:2020wlx}), but is beyond the scope of this work. 

\subsubsection{Resonant pion production}
A non-negligible fraction of CC resonant production (CCRES) contributes to the CC0$\pi$ topology, which occurs when the outgoing pion is absorbed inside the nucleus through FSI. This is particularly important for multi-GeV neutrinos like in the \minerva flux, leading the CCRES contribution to be larger than in T2K. To account for this, a normalization parameter that varies the amount of CCRES events with an absorbed pion in this specific topology is considered. Since the impact of FSI is nuclear target dependent, this parameter is split for interactions on carbon and oxygen targets. 

In addition, we further consider three parameters that modify the Rein--Sehgal model with lepton mass effects and updated form factors~\cite{Rein:1980wg,Rein:1981ys,Graczyk:2007xk,Graczyk:2014dpa,Berger:2007rq} implemented in \texttt{NEUT}. The parameters of this model are: the axial mass, $M_A^\mathrm{RES}$, the value of the axial form factor when $Q^2 = 0$, $C_5^A$, and the normalization of the non-resonant $1/2$-isospin background, $I_{1/2}$. Their prior values and uncertainties are fixed from recently-updated fits to bubble chamber data on hydrogen and deuterium from ANL~\cite{PhysRevD.25.1161} and BNL~\cite{PhysRevD.34.2554}, which has been used in Ref.~\cite{t2kOA2022_neutrino2022, t2koa2022wip}.

\subsubsection{2p2h interactions}
As already mentioned, 2p2h interactions also populate the CC0$\pi$ topology. \texttt{NEUT} describes 2p2h interactions using the Nieves \etal model~\cite{Nieves:2011yp}, whose cross section features two distinct peaks in the energy and momentum transfer that broadly correspond to $\Delta$ and non-$\Delta$ excitations. 

An uncertainty on the amount of 2p2h events is added as a simple normalization parameter which is able to adjust the number of 2p2h interactions for each target. For this parameter, we opt for a 30\% prior uncertainty to cover differences between the \texttt{NEUT} model and the alternative 2p2h calculation from the SuSAv2 theory group~\cite{Megias:2016lke,Dolan:2019bxf}. Additionally, we prescribe a 2p2h shape uncertainty which varies the relative strength between the $\Delta$ and non-$\Delta$ contributions, as used and described in Refs.~\cite{T2K:2023smv,T2K:2017rgv,T2K:2021xwb}.


\bgroup
\def\arraystretch{1.3}
\begin{table*}[]
    \centering
    \begin{tabular}{|l|c|c|c|}
        \hline 
        \textbf{Parameter} & \textbf{Central value} & \textbf{Prior uncertainty ($1\sigma$)} & \textbf{Notes} \\
        \hline
        \multicolumn{4}{l}{\textbf{Carbon parameters}} \\
        \hline
        {$p$-shell norm.\ C} & 0 & 20\% & \multirow{2}{*}{$1\sigma$ variation changes CCQE cross section by 10\%}\\ 
        \cline{2-3}
        {$s$-shell norm.\ C} & 0 & 40\% &  \\ 
        \hline
        {$p$-shell shape C} & 0 & 100\% & \multirow{2}{*}{From $(e,e'p)$ data, used only for fits in \dpt (CH target)}\\ 
        \cline{2-3}
        {$s$-shell shape C} & 0 & 100\% & \\ 
        \hline 
        {SRC norm.\ C} & 1 & 100\% & \\ 
        \hline 
        {Pauli Blocking C (MeV/$c$)} & 209 & 30 & \multirow{2}{*}{Used only with fits in lepton kinematics} \\ 
        \cline{1-3}
        {Optical Potential C} & 50\% & 50\% & \\ 
        \hline 
        {2p2h norm.\ C} & 1 & 30\% & \\ 
        \hline 
        {2p2h shape C} & 0 & 300\% &  Defined in Ref.~\cite{T2K:2021xwb}\\ 
        \hline 
        {Pion abs.\ norm.\ C} & 1 & 30\% & \\ 
        \hline
        \multicolumn{4}{l}{\textbf{Oxygen parameters}} \\
        \hline
        {$p_{1/2}$-shell norm.\ O} & 0 & 45\% & \multirow{3}{*}{$1\sigma$ variation changes CCQE cross section by 10\%}\\ 
        \cline{2-3}
        {$p_{3/2}$-shell norm.\ O} & 0 & 25\% & \\ 
        \cline{2-3}
        {$s$-shell norm.\ O} & 0 & 75\% & \\ 
        \hline 
        {SRC norm.\ O} & 1 & 100\% & \\ 
        \hline 
        {Pauli Blocking O (MeV/$c$)} & 209 & 30 & \multirow{2}{*}{Used only with fits in lepton kinematics} \\ 
        \cline{1-3}
        {Optical Potential O} & 50\% & 50\% & \\ 
        \hline 
        {2p2h norm.\ O} & 1 & 30\% & \\ 
        \hline 
        {2p2h shape O} & 0 & 300\% & Defined in Ref.~\cite{T2K:2021xwb}\\ 
        \hline 
        {Pion abs.\ norm.\ O} & 1 & 30\% & \\ 
        \hline
        \multicolumn{4}{l}{\textbf{Nucleon interaction parameters}} \\
        \hline
        $M_A^\mathrm{QE}$ (GeV/$c^2$) & 1.03 & 0.06 & Dipole parametrization \\ 
        \hline 
        {High $Q^2$ norm.\ 1} & 1 & 11\% & $Q^2 \in [0.25, 0.50 \text{ GeV}^2[$\\ 
        \cline{2-4} 
        {High $Q^2$ norm.\ 2} & 1 & 18\% & $Q^2 \in [0.50, 1.00 \text{ GeV}^2[$\\ 
        \cline{2-4} 
        {High $Q^2$ norm.\ 3} & 1 & 40\% & $Q^2 \in [1.00 \text{ GeV}^2, +\infty[$\\ 
        \hline 
        $M_A^\mathrm{RES}$ (GeV/$c^2$) & 0.91 & 0.1 & \multirow{3}{*}{Correlated~\cite{t2kOA2022_neutrino2022, t2koa2022wip}}  \\  
        \cline{2-3}
        {$C_5^A (Q^2=0)$} & 1.06 & $10\%$ & \\ 
        \cline{2-3}
        {$I_{1/2}$ non-res.\ bkg.} & 1.21 & $27\%$ & \\ 
        \hline
        \multicolumn{4}{l}{\textbf{FSI parameters}} \\
        \hline
        {With nucleon FSI} & 1 & 30\% & \multirow{2}{*}{\shortstack{Strongly anti-correlated, see Sec.~\ref{sec:FSI} \\ Used only in fits to semi-inclusive measurements\footnote{I.e.\ measurements with restrictions on the outgoing hadron kinematics, corresponding to the three fits of Sec.~\ref{sec:fit0pNp} and the fit of Sec.~\ref{sec:fitMNV}}}}\\ 
        \cline{2-3}
        {Without nucleon FSI} & 1 & 30\% & \\ 
        \hline 
    \end{tabular}
    \caption{Summary of the parameters introduced in Sec.~\ref{sec:syst} and their prior uncertainties.}
    \label{tab:priors}
\end{table*}
\egroup

\section{Fits to CC$0\pi$ cross-section measurements}
\label{sec:fit}

Fits to a variety of available neutrino cross-section measurements were performed with the goal of benchmarking the new proposed parametrization; understanding the extent to which these systematic uncertainties can be constrained by data; and assessing the compatibility of these constraints with expectations from electron scattering data. 
The topology of interest is CC0$\pi$, since it is the most sensitive to the CCQE contribution and thus to this new set of parameters. Since most of these parameters can alter both the lepton and nucleon kinematics for CCQE interactions consistently---with the exception of OP and FSI parameters---we fit to cross-section measurements reported in muon kinematics and the transverse kinematic imbalance between the outgoing lepton and nucleons, such as \dpt. In addition, since the parameters have been developed both for carbon and oxygen targets, it is particularly interesting to validate them against available data for both elements. For these reasons, the following measurements have been chosen for the fits: the T2K CC0$\pi$ cross section in muon kinematics $(p_\mu,\cos\theta_\mu)$ on oxygen and carbon~\cite{T2K:OeC}, the T2K CC0$\pi$ cross section in \dpt, muon, and proton kinematics on CH~\cite{T2K:2018rnz}, and the \minerva CC0$\pi$ cross section in \dpt on CH~\cite{MINERvA:2018hba}. The oxygen and carbon measurement from T2K is a simultaneous measurement of the cross section on both targets, and as such provides uncertainties correlated across them.

Each cross-section measurement is not sensitive to all the parametrized uncertainties introduced in Sec.~\ref{sec:syst}, and as such not every fit includes all the parameters. Tab.~\ref{tab:priors} states when parameters are only included in a sub-set of the fits. The oxygen uncertainties are only fit when considering the T2K measurement that includes an oxygen target. The parameters that affect the low energy transfer region (OP and PB) are only included in cross-section measurements that are sensitive to low energy transfer interactions (i.e.\ not those that require the observation of a proton in the final state -- such that its momentum must be above the 450 MeV/$c$ or 500 MeV/$c$ experimental tracking thresholds (the exact value depends on which measurements are being considered) -- as this implies the need for a sufficiently high energy transfer that the implemented corrections to PWIA are not relevant). 
The parameters that affect the FSI intranuclear cascade are only included in fits sensitive to outgoing nucleon kinematics (i.e.\ those sensitive to the semi-inclusive CCQE cross section). 

Fits are performed using the \texttt{NUISANCE} framework~\cite{Stowell:2016jfr}. Each of the cross-section measurements used in this work were already implemented in \texttt{NUISANCE}, including the bin-to-bin covariance matrices as published by the experiments. The minimization package used in the fits is MINUIT~\cite{James:1975dr}.

\subsection{Chi-squared test-statistic}
To perform the desired fits, we developed and implemented in \texttt{NUISANCE} a specific $\chi^2$ test-statistic, with the explicit goal of mitigating the impact of Peelle's Pertinent Puzzle (PPP)~\cite{DAgostini:1993arp,ppp}, that is well known to cause a preference for artificially low normalization fit results when a good fit to highly correlated data cannot be found. 

Usually, a fit can be made by comparing varied simulated predictions to a cross-section measurement using the measurement histogram $B = \{B_1, ..., B_n\}$ and covariance matrix $M=\mathrm{Cov}\left[\{B_i\}\right]$ as provided by the experiments and minimizing a standard test-statistic like the chi-squared:
\begin{equation}\label{eq:chi2}
    \chi_\mathrm{}^2 = \sum_{i,j} \left(B_i - B_i^\mathrm{MC}\right) \left( M^{-1} \right)_{i,j} \left(B_j - B_j^\mathrm{MC}\right)
\end{equation}
where $B^\mathrm{MC}=(B^\mathrm{MC}_1, ..., B^\mathrm{MC}_n)$ corresponds to the bins in the histogram of simulated predictions. When the bins in the measurement are highly correlated, fits can converge to simulated distributions with an unphysical low overall normalization when a good fit is not possible, which is known as PPP. Whilst PPP is ultimately just a consequence of correlations in the covariance matrix, the preference for low normalizations partially stems from the $\chi^2$ in Eq.~\ref{eq:chi2} assuming that the absolute uncertainty on each bin of the measurement is independent of its normalization. This implies that the relative uncertainty is larger when fitting to models that predict lower normalizations. When a fitted parametrization cannot match the shape of a measured cross section, the $\chi_\mathrm{}^2$ will therefore tend to be less poor at low normalizations.

If the Gaussian assumption implicit in the standard covariance matrix is correct, then this preference for models with low normalizations is genuine and there is no puzzle (beyond why the fitted parametrization is insufficient). However, it is expected that some types of experimental uncertainty that control the overall normalization of the measured cross section (for example flux or background subtraction systematic uncertainties) should not follow a Gaussian, but rather a log-Gaussian distribution~\cite{DAgostini:1993arp}. In the case of non-Gaussian uncertainties the $\chi^2$ in Eq.~\ref{eq:chi2} may not represent a good test statistic in a fit.

Methods to work around PPP have been employed, such as using a shape-only chi-square~\cite{T2K:2020jav,Dolan:2018zye}, or neglecting the bin-to-bin correlations~\cite{MicroBooNE:2021ccs}. However, neither of these methods are satisfactory since the former ignores the valuable information on the total cross section from the measurement, while the latter is almost meaningless when the measured cross section has significant correlations between bins (as most do). We propose an alternative way to mitigate this effect which consists of separating the normalization and shape contributions of the covariance matrix, that corresponds to isolating the relative uncertainty from the uncertainty on the overall cross-section normalization, in such a way to make the absolute uncertainty larger for models predicting lower normalizations. This method results in a relative uncertainty that remains constant as a function of the normalization, as motivated by the arguments of Ref.~\cite{DAgostini:1993arp}. Such a treatment is generally well motivated for data dominated by correlations due to multiplicative uncertainties, such as the overall normalization uncertainties from flux uncertainties that often dominate cross-section measurements. 

Concretely, this can be obtained by applying a transformation to both the data and simulated histograms as well as to the covariance matrix, in order to separate them into a ``shape'' and a ``norm'' part. The new histograms $C = \{C_1, ..., C_n\}$ are defined as:
\begin{equation}
C_i = f(B_i) =
  \begin{cases}
    \alpha \frac{B_i}{\sum_k B_k}, & 1 \leq i \leq n-1 \\
    B_T = \sum_k B_k, & i=n
  \end{cases}
\end{equation}
where $\alpha$ is a scale parameter. Note that if $B$ is a differential rather than absolute number of events, scaling by the bin width is required. 
The function $f: B \mapsto C$ is bijective, meaning that no information is lost by moving from $B$ to $C$.

In this new basis, a covariance matrix, $N=\mathrm{Cov}\left[\{C_i\}\right]$, would ideally be built using the same ``toys'' or ``universes'' experiments typically use to build their standard covariance. Unfortunately such information is not usually provided in experimental data releases. Instead, the covariance matrix is approximated via a non-linear transformation of the original covariance matrix $M$, using the following formula:  
\begin{equation}
N = J(f).M.J(f)^T
\end{equation}
where $J(f)$ is the Jacobian of the non-linear transformation $f$.
The new covariance matrix is expressed as follows: 
\begin{widetext}
\begin{equation}
   N = 
    \begin{pmatrix}
        (N_\mathrm{S})_{i,j} = \frac{\alpha^2}{B_T^2} \left( M_{i,j} - \frac{B_i}{B_T} \sum_l M_{i,l} - \frac{B_j}{B_T} \sum_k M_{k,j} - \frac{B_i B_j}{B_T^2} \sum_{kl} M_{k,l} \right)& \rvline &
        \begin{matrix}
            \frac{\alpha}{B_T} \left(\sum_l M_{1,l} - \frac{B_1}{B_T} \sum_{kl} M_{k,l} \right)\\[1em]
            \vdots\\[1em]
            \frac{\alpha}{B_T} \left(\sum_l M_{n-1,l} - \frac{B_{n-1}}{B_T} \sum_{kl} M_{k,l} \right)
        \end{matrix}\\[1em]
        \hline
        \begin{matrix}
            \frac{\alpha}{B_T} \left(\sum_k M_{k,1} - \frac{B_1}{B_T} \sum_{kl} M_{k,l} \right) &
            \cdots &
            \frac{\alpha }{B_T} \left(\sum_k M_{k,n-1} - \frac{B_{n-1}}{B_T} \sum_{kl} M_{k,l} \right)
        \end{matrix}
        & \rvline & \sum_{kl} M_{k,l}
    \end{pmatrix} 
\end{equation}
\end{widetext}
The matrix $N$ has the same dimension and the same positive-definiteness properties as $M$ since the mapping $B \mapsto C$ is a bijection. 
$N$ is composed of two diagonal blocks: the $N_\mathrm{S}$ block which corresponds to the shape-only covariance, and the $\sum_{kl} M_{k,l}$ element which corresponds to the data norm variance. The off-diagonal blocks represent the correlations between the norm and the shape components. 

Finally, by transforming the MC histogram using this same function $f$, the ``norm-shape" (NS) chi-square can be computed in this basis as:
\begin{equation}
    \chi_\mathrm{NS}^2 = \sum_{1 \leq i,j \leq n} \left(C_i - C_i^\mathrm{MC}\right) \left( N^{-1} \right)_{i,j} \left(C_j - C_j^\mathrm{MC}\right).
    \label{eq:NSchi2}
\end{equation}
This new computation of the covariance matrix and the chi-square was implemented in \texttt{NUISANCE} and the correctness of the implementation was validated by comparisons with covariance matrices computed from toys. The use of the $\chi_\mathrm{NS}^2$ allows to mitigate the PPP problem observed when using the standard $\chi^2$ of Eq.~\ref{eq:chi2} in the fit. This method was first discussed in Ref.~\cite{Chakrani:2022tey,Koch:2021yda} and also used, via our implementation, in Refs.~\cite{GENIE:2022qrc,MicroBooNE:2021ccs}\footnote{On a historical note: the MiniBooNE collaboration had developed (but not published) a similar approach in the context of decomposing covariance matrices into total, shape-only and ``mixed'' terms. This is noted on page 217 of Ref.~\cite{Katori:2008zz}.}. 

It should be noted that whilst this method mitigates the impact of PPP, it also makes assumptions that: 
\begin{itemize}
    \item the real uncertainty in the data did not follow the multivariate Gaussian covariance reported by the experiments, but rather follows a distribution from which the relative uncertainty is constant as a function of the measurement's normalization, 
    \item the transformation of the experimentally reported covariance provides a better description of the real distribution of the measurement uncertainties. 
\end{itemize}
Given the limited information provided in experimental data releases, it is not possible to test these assumptions\footnote{If experiments were to provide a distribution of cross-section results mapped by their uncertainties (i.e.\ the ``universes'' used to calculate the covariance matrix that is usually supplied) it may be possible to tailor test statistics used in fits to best describe the plausible variations of the measured cross section.}. As such, the modified test-statistic is used for fits but the usual $\chi^2$ from Eq.~\ref{eq:chi2} is also reported.

\subsection{Fit results}\label{sec:fitresults}

In this section, the results of the fits, using the $\chi_\mathrm{NS}^2$ defined in the previous section, to T2K and \minerva data are presented and discussed.


\subsubsection{Fit to T2K CC0$\pi$ cross section data on oxygen and carbon in muon kinematics}\label{sec:fitOC}

\begin{figure*}[]
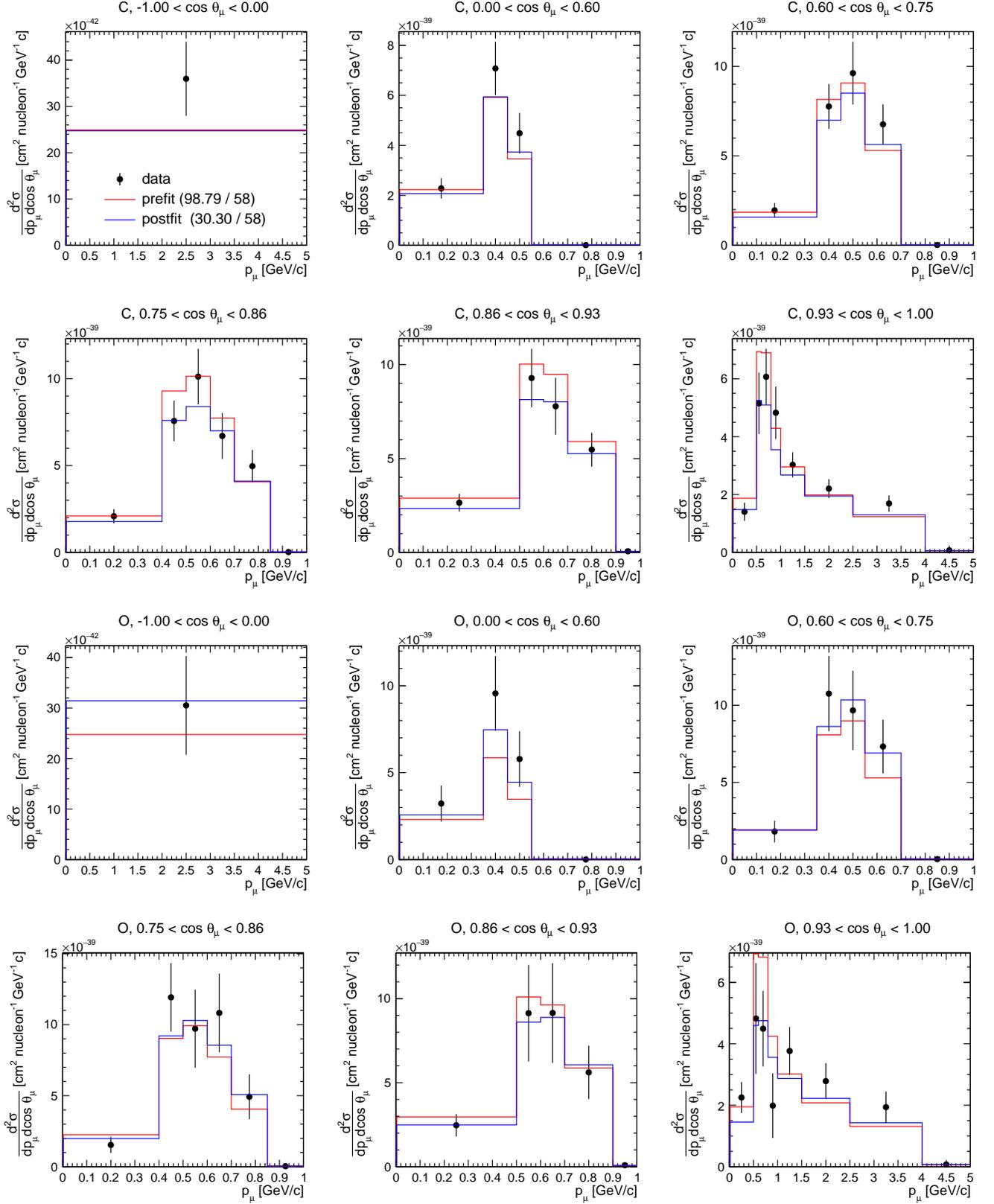

    \centering
    \foreach \a in {1,...,6}
    {
        \includegraphics[page=\a, width=0.32\textwidth]{plots/FitResults/T2K_OC/pmu_slices_C.pdf}
    }
    \foreach \a in {1,...,6}
    {
        \includegraphics[page=\a, width=0.32\textwidth]{plots/FitResults/T2K_OC/pmu_slices_O.pdf}
    }

    \caption{Prefit (red) and postfit (blue) distributions of $p_\mu$ in bins of $\cos \theta_\mu $ from the fit to T2K CC0$\pi$ joint measurement of lepton kinematics on carbon and oxygen. The usual chi-squares as well as the number of bins are quoted in the legend. The NS chi-square $\chi^2_\mathrm{NS}$ used in the minimization is reported in Tab.~\ref{tab:chi2}.}
    \label{fig:t2k_OC}

\end{figure*}

\begin{figure*}[]
    \centering
    \includegraphics[width=0.85\linewidth]{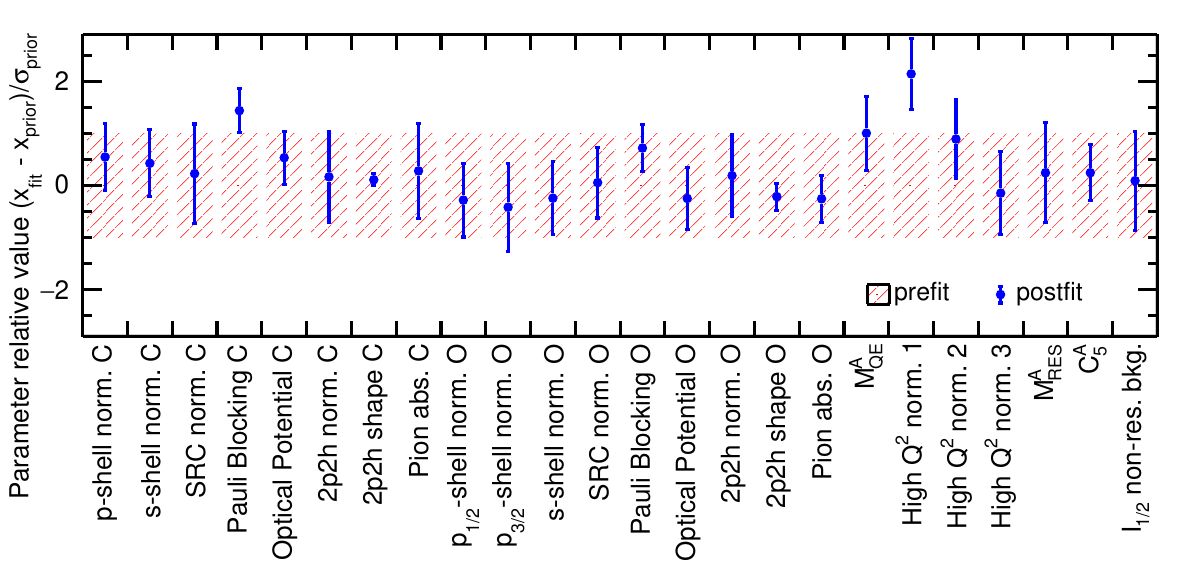}
    \caption{Prefit (red) and postfit (blue) values and constraints on the uncertainties from the fit to T2K CC0$\pi$ joint measurement of lepton kinematics on carbon and oxygen. The displayed central value for each parameter corresponds to the difference with respect to its prior value divided by the prior uncertainty as reported in Tab.~\ref{tab:priors}. }
    \label{fig:t2k_oc_params}
\end{figure*}
\begin{figure*}[]
    \centering
    \includegraphics[width=0.85\linewidth]{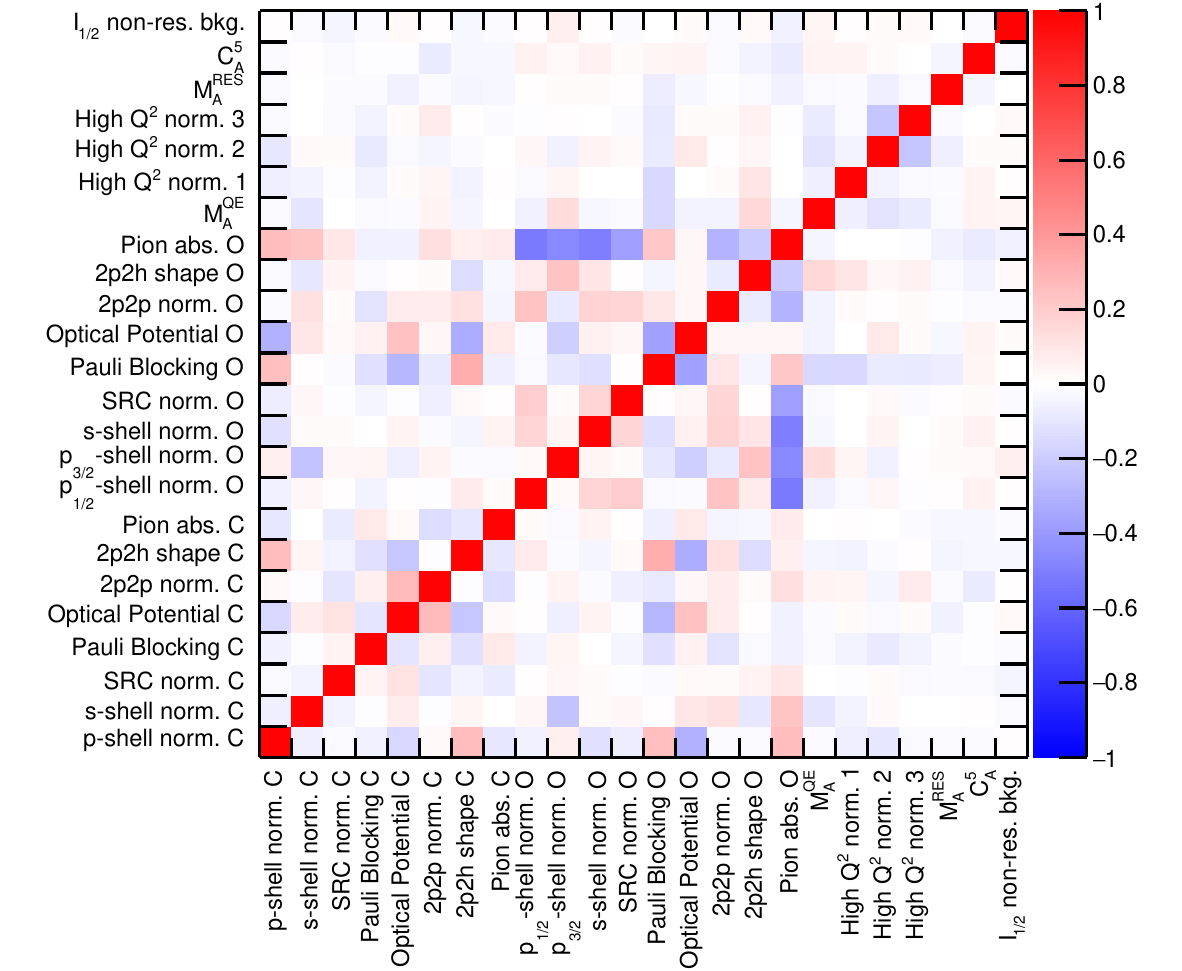}
    \caption{Postfit correlation matrix from the fit to T2K CC0$\pi$ joint measurement of lepton kinematics on carbon and oxygen.}
    \label{fig:t2k_oc_corr}
\end{figure*}

Fig.~\ref{fig:t2k_OC} shows the prefit and postfit cross-section predictions for the T2K CC0$\pi$ measurement on carbon and oxygen targets from Ref.~\cite{T2K:OeC}, whilst Fig.~\ref{fig:t2k_oc_params} and Fig.~\ref{fig:t2k_oc_corr} show the prefit and postfit parameter values and the associated correlation matrix. The prefit, i.e.\ the nominal spectra predicted by \texttt{NEUT}, and the postfit distributions are displayed and compared with the data and the value of the corresponding $\chi_\mathrm{NS}^2$ as well as the usual $\chi^2$ are reported in Tab.~\ref{tab:chi2}. 

\begin{center}
\begin{table*}[]
    \begin{tabular}{|l||c|c||c|c||c|}
    \hline
        \textbf{Measurement} & {Prefit $\chi^2_\mathrm{NS}$} & {Postfit $\chi^2_\mathrm{NS}$} & Prefit $\chi^2$ & Postfit $\chi^2$ & Number of bins \\
        \hline \hline
        {T2K oxygen + carbon (Sec.~\ref{sec:fitOC})} & 110.88 & 35.81 & 98.79 & 30.30  & 58\\ 
        \hline \hline
        {T2K CC$0\pi Np$ $\dpt$ only (Sec.~\ref{sec:fit0pNp})} & 12.59 & 7.37 & 15.72 & 8.48 & 8\\ 
        \hline 
        {T2K CC$0\pi 0p$ $(p_\mu, \cos \theta_\mu)$ only (Sec.~\ref{sec:fit0pNp})} & 144.35 & 75.13 & 107.57 & 62.55 & 50\\ 
        \hline 
        {T2K CC$0\pi 0p$ + CC$0\pi Np$ (Sec.~\ref{sec:fit0pNp})} & $144.35 + 14.56$ & $86.80 + 10.01$ & $107.57 + 16.76$ & $64.19 + 11.83$ & $50 + 8$ \\ 
        \hline \hline
        {\minerva $\dpt$ (Sec.~\ref{sec:fitMNV})} & 109.10 & 79.51 & 114.32 & 76.14 & 24 \\ 
        \hline
    \end{tabular}
\caption{Summary of the prefit and postfit norm-shape chi-square $\chi^2_\mathrm{NS}$ used in the minimization, and the usual chi-square $\chi^2$ for reference in the different fits presented in Sec.~\ref{sec:fitresults} along with the corresponding number of bins. }
\label{tab:chi2}
\end{table*}
\end{center}

It can immediately be noted that the data/model agreement is dramatically improved after the fit adjustment of the systematic parameters, which is also reflected in the important decrease in both $\chi_\mathrm{NS}^2$ and $\chi^2$. Importantly, both are also lower than the number of bins, implying a quantitatively good agreement between the post-fit model and the measured cross section. It should be noticed that, as detailed in Ref.~\cite{T2K:OeC}, the prefit data/model disagreement comes mainly from the bins that correspond to forward lepton kinematics (i.e.\ the most forward cos$\theta_\mu$ slice). 
This is the region that corresponds to a low energy transfer which is known to be more complicated to model due to the effects beyond the PWIA. For instance RFG and LFG models are often corrected using the random phase approximation to achieve better agreement in this region of kinematic phase space~\cite{Nieves:2011pp}. It is thus expected to see that the postfit agreement in this region is driven by the parametrization of physics beyond PWIA discussed in Sec.~\ref{sec:pwiauncert}: PB and OP. Indeed, Fig.~\ref{fig:t2k_oc_params} shows the preference for an application of the OP correction and more PB than is in the nominal \texttt{NEUT}. Interestingly, the strength of these corrections to PWIA appear to be stronger on carbon than on oxygen. 

Fig.~\ref{fig:t2k_oc_params} shows that the measurement offers fair sensitivity to constraining the shell normalization parameters, which assist in compensating the effects of the PB and OP parameters. The overall CCQE normalization and its shape in $Q^2$ is also adjusted by a significant variation of the $M_A^\mathrm{QE}$ and high $Q^2$ parameters. 
Almost all parameters remain reasonable given their prior expectations both as encoded in the prefit uncertainties and as would be expected from electron scattering data. In particular, the shell normalizations are not pulled far from their nominal values.
However, it can be noted that PB for carbon is pulled outside of its relatively conservative prior value to give an effective Fermi momentum of $\sim$250 MeV/$c$. Interpreted as a Fermi-gas Fermi momentum, this would imply nonphysical nuclear density~\cite{Alvarez-Ruso:2023ovr} and so this is may be indicative that PB is acting partially as an effective parameter to account for missing freedom within the model (e.g.\ to span differences between different approaches to PB or other physics beyond PWIA).

The postfit correlation matrix is shown in Fig.~\ref{fig:t2k_oc_corr}. There are anticorrelations between the PB and OP uncertainties for carbon and more prominently for oxygen since both parameters have a similar impact on the cross section. Anticorrelations are also been between the pion absorption parameter and the parameters that control the CCQE normalization, indicating that the data is not able to offer isolated constraints on the QE and non-QE model components (which is to be expected when only measuring outgoing lepton kinematics). 

It may be noteworthy to highlight that the parametrization developed in this work treats the oxygen and carbon uncertainties as two independent groups (except for the nucleon-level parameters such as $M_A^\mathrm{QE}$, $M_A^\mathrm{RES}$, etc., which are common), but the postfit correlation matrix shown in Fig.~\ref{fig:t2k_oc_corr} exhibits correlations between these two sets of parameters. As previously stated, the measurement corresponds to a \textit{joint} cross section on the two targets, the data covariance includes correlations between carbon bins and oxygen bins, which is then reflected on the parameters. Such simultaneous measurements are increasingly important for oscillation measurements to better understand the extrapolation of model constraints between targets.

\subsubsection{Fit to T2K cross section data in CC0$\pi$0p and CC0$\pi N$p topologies on hydrocarbon}
\label{sec:fit0pNp}

\begin{figure}[]
    \centering
    \includegraphics[width=0.95\linewidth]{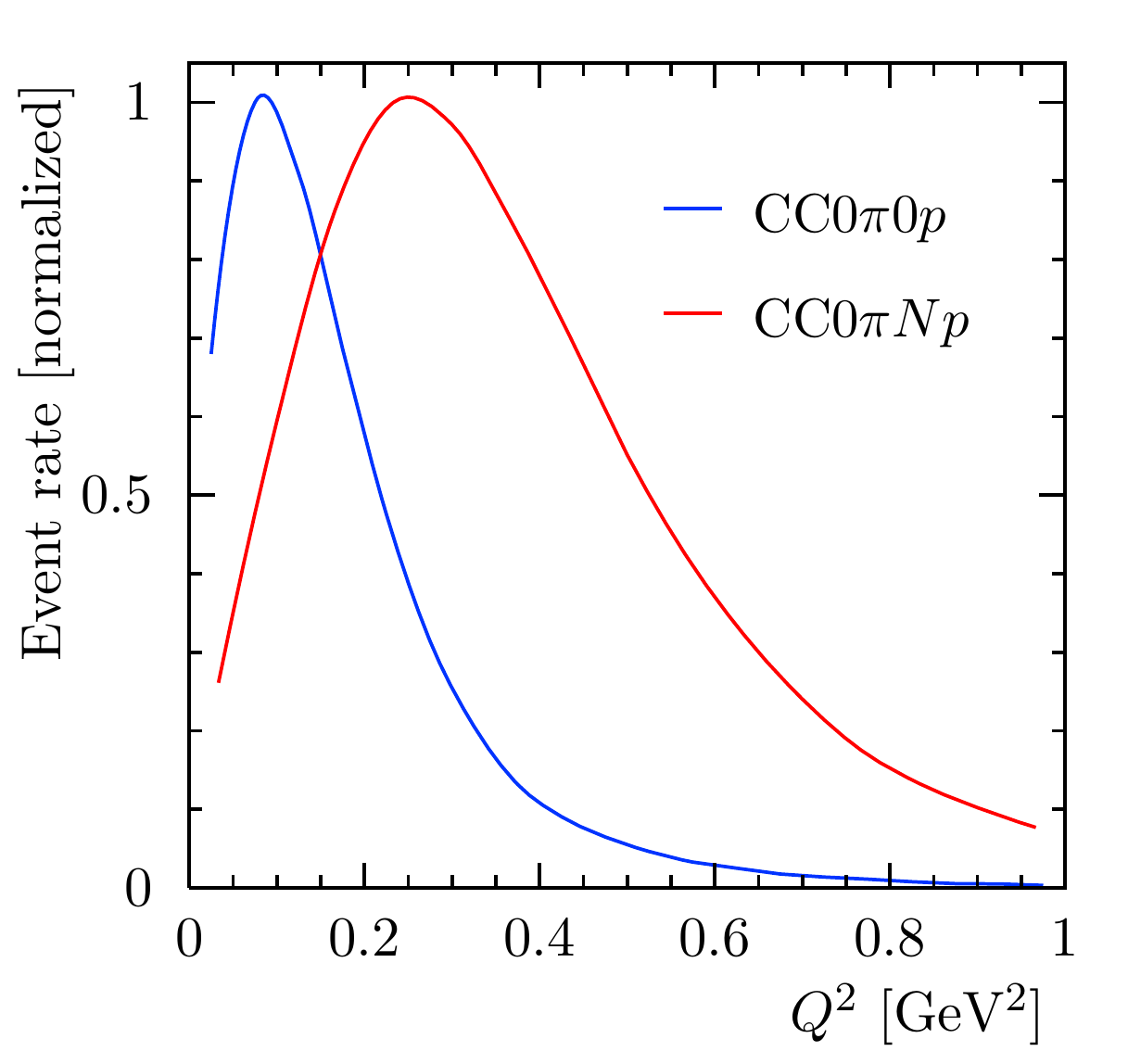}
    \includegraphics[width=0.95\linewidth]{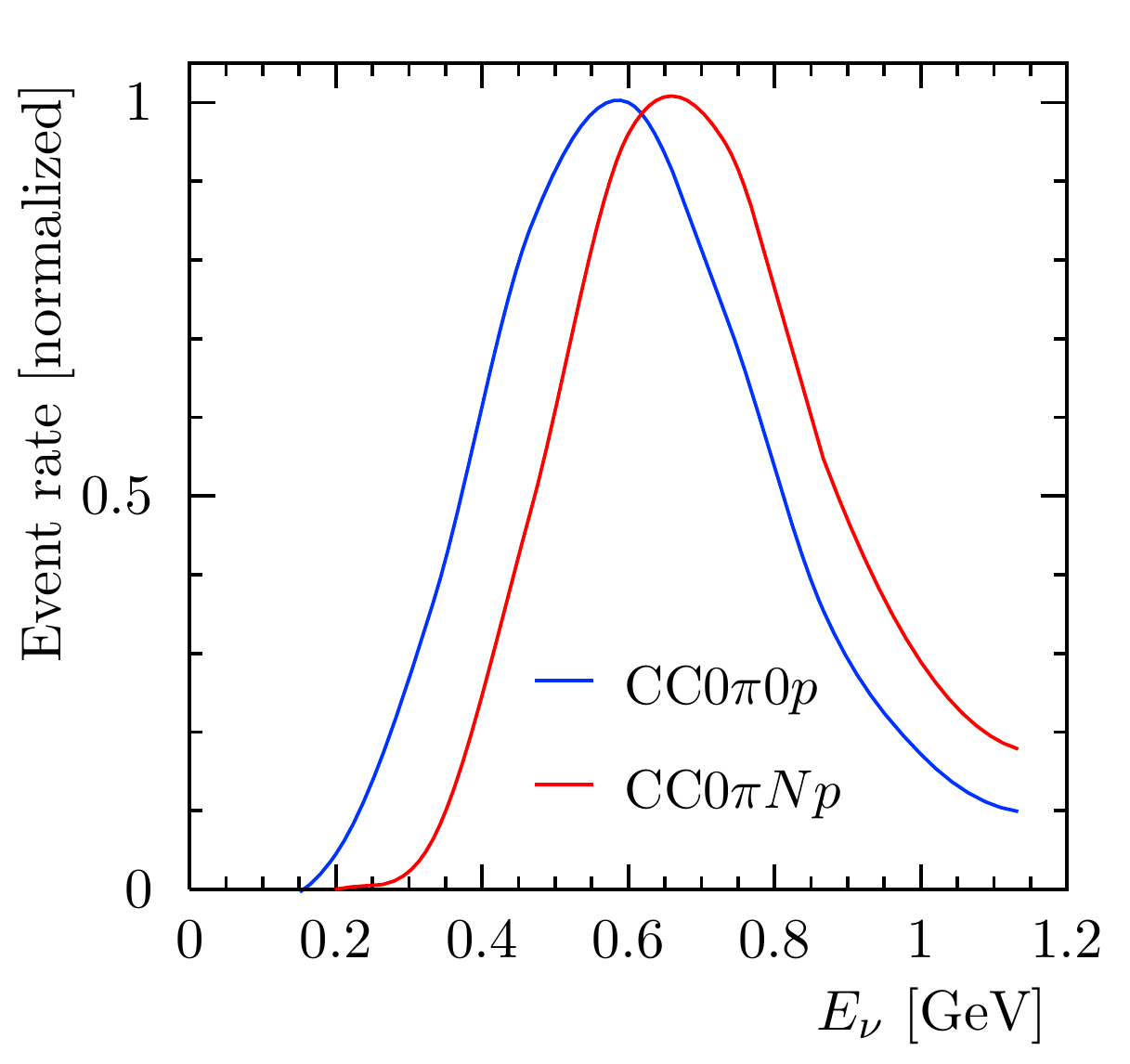}
    \caption{True squared four momentum transfer (upper) and neutrino energy (lower) distributions of interactions falling into the the CC$0\pi 0p$ (blue) and CC$0\pi Np$ (red) topologies in the T2K beam on a carbon target as predicted by \texttt{NEUT}.}
    \label{fig:enu_0pNp}
\end{figure}

In this section, we fit T2K CC$0\pi$ cross-section measurements on hydrocarbon as a function of transverse momentum imbalance $\dpt$ and final-state muon kinematics for CC$0\pi$ topologies with (CC$0\pi Np$, $N \geq 1$) and without (CC$0\pi 0p$) a measured proton in the final state respectively~\cite{T2K:2018rnz}. Note that $0p$ and $Np$ refer to protons above and below tracking threshold ($\sim$500~MeV/c). Three distinct fits are performed:
\begin{enumerate}[label=(\alph*)]
    \item fitting lepton kinematics in CC$0\pi 0p$,
    \item fitting $\dpt$ in CC$0\pi Np$,
    \item simultaneously fitting lepton kinematics in CC$0\pi 0p$ and $\dpt$ in CC$0\pi Np$.
\end{enumerate}
One of the interests of performing this simultaneous fit is to evaluate the ability of the model to describe neutrino interactions in different neutrino energy ranges and in different regions of lepton kinematics. Indeed, the CC$0\pi 0p$ and the CC$0\pi Np$ topologies correspond to distinct regions of momentum transfer (as a proton requires $\sim$450~MeV momentum to be observed) and therefore to different regions of $E_\nu$, as illustrated in Fig.~\ref{fig:enu_0pNp}. It also tests the parametrization's ability to describe how lepton kinematics changes as a function of the outgoing proton kinematics. It should be noted that the two measurements considered in (c) are treated as uncorrelated (as correlations between the two measurements are not available) despite not being completely independent. The kinematic overlap between them is expected to be very small, given that CC$0\pi 0p$ topology allows protons up to 500~MeV/$c$, whilst the $\dpt$ measurement considers only protons above 450~MeV/$c$ although a more complete analysis would provide correlations from systematic uncertainties. 

In our analysis the highest outgoing muon momentum bin (which extends up to 30~GeV/$c$) in each angular slice of the lepton kinematics measurement in CC$0\pi 0p$ is removed. These constitute a negligible fraction of the total measured cross section and are in a momentum range where T2K cannot guarantee reliable reconstruction\footnote{Whilst these bins are in the data release, they were not shown within Ref.~\cite{T2K:2018rnz}.}.

\begin{figure}[]
    \centering
    \includegraphics[width=0.95\linewidth]{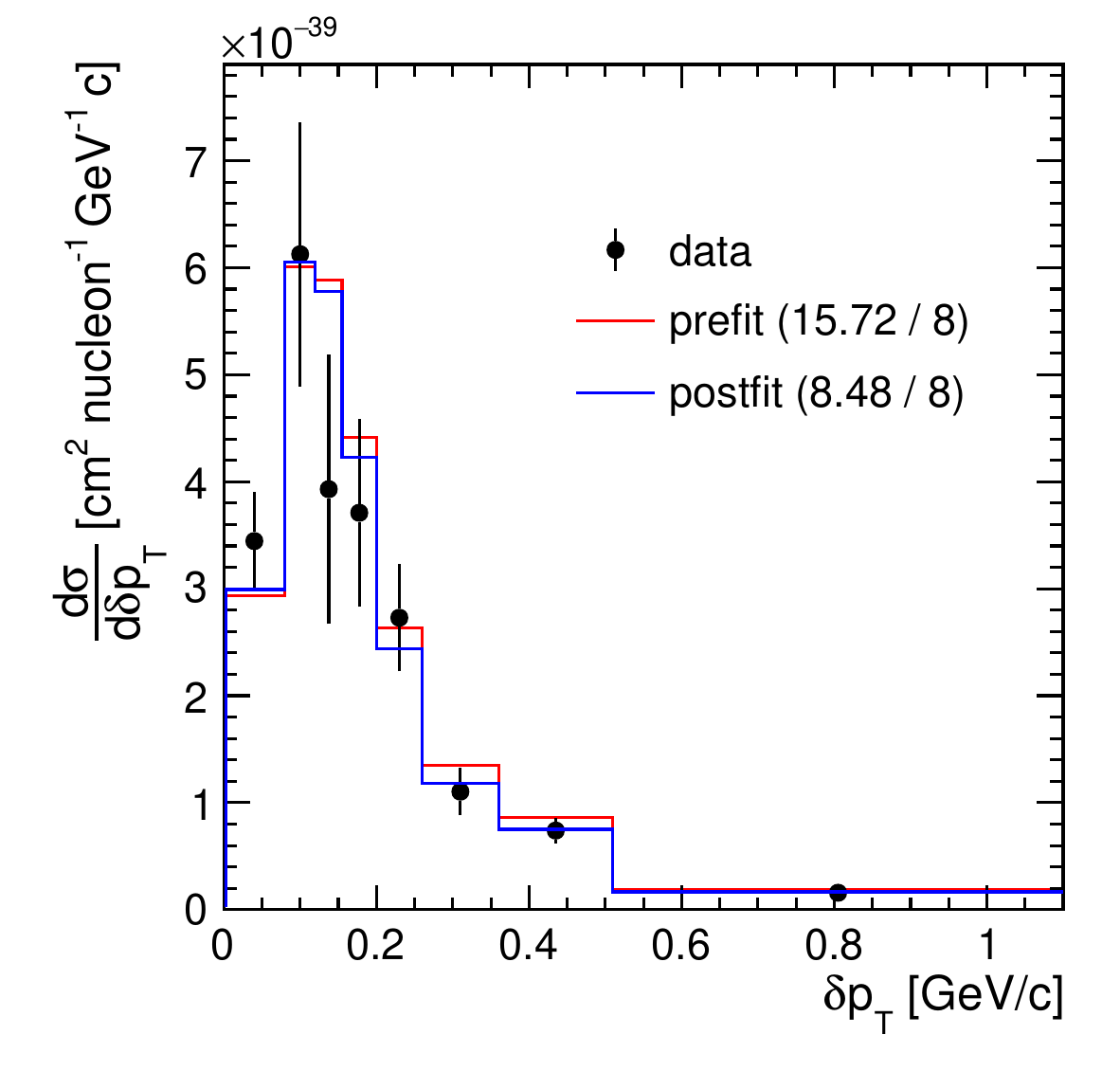}
    \caption{Prefit (red) and postfit (blue) distributions of $\dpt$ from fitting the T2K CC0$\pi Np$ data only. The usual chi-squares as well as the number of bins are quoted in the legend. The NS chi-square $\chi^2_\mathrm{NS}$ used in the minimization is reported in Tab.~\ref{tab:chi2}.}
    \label{fig:t2k0pNp_dpt}
\end{figure}

\begin{figure*}[]
    \centering
    \includegraphics[width=0.7\linewidth]{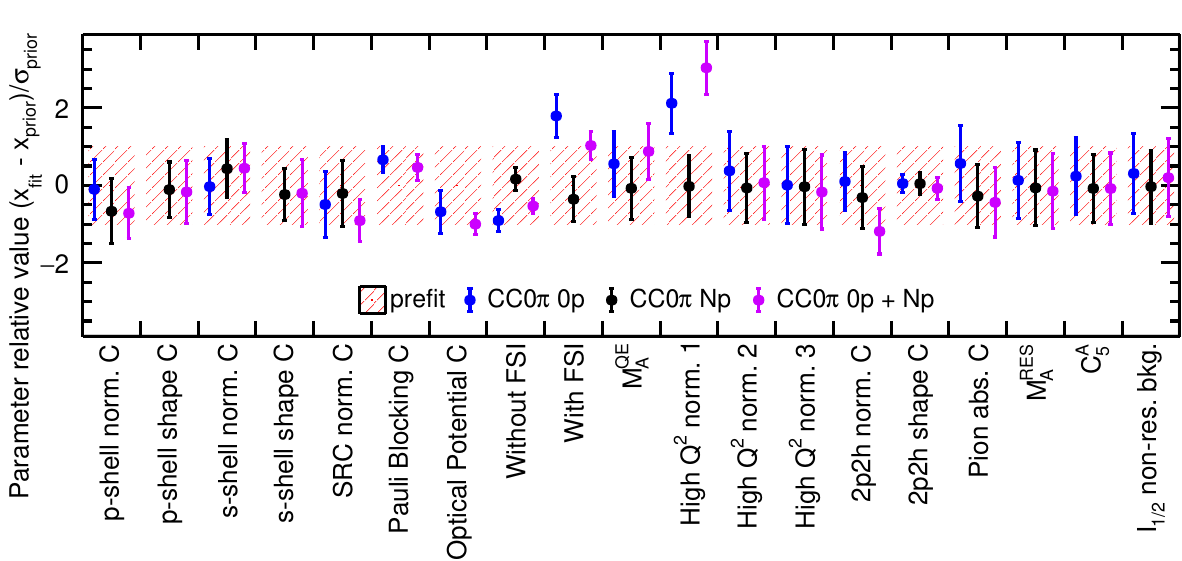} 
    \caption{Prefit and postfit values and constraints on the uncertainties from the fit to T2K CC0$\pi 0p$ measurement of lepton kinematics and CC0$\pi Np$ measurement of $\dpt$ on carbon. The displayed central value for each parameter corresponds to the difference with respect to its prior value divided by the prior uncertainty as reported in Tab.~\ref{tab:priors}.}
    \label{fig:t2k0pNp_params}
\end{figure*}

\begin{figure*}[]
    \centering
    \includegraphics[page=3, width=0.495\linewidth]{plots/FitResults/T2K_0pNp/postfitcov_0pNp.pdf}
    \includegraphics[width=0.495\linewidth]{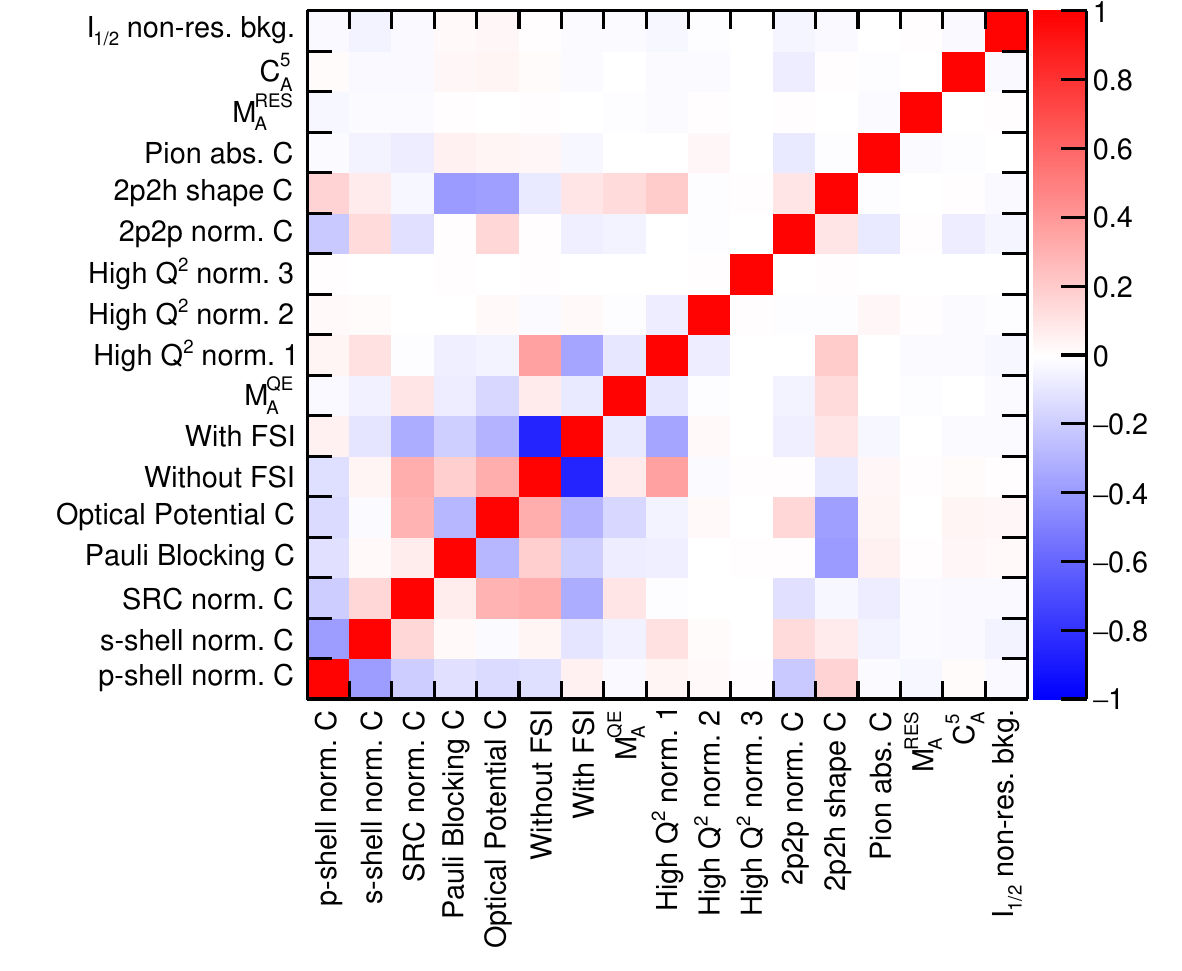} 
    \includegraphics[page=1, width=0.495\linewidth]{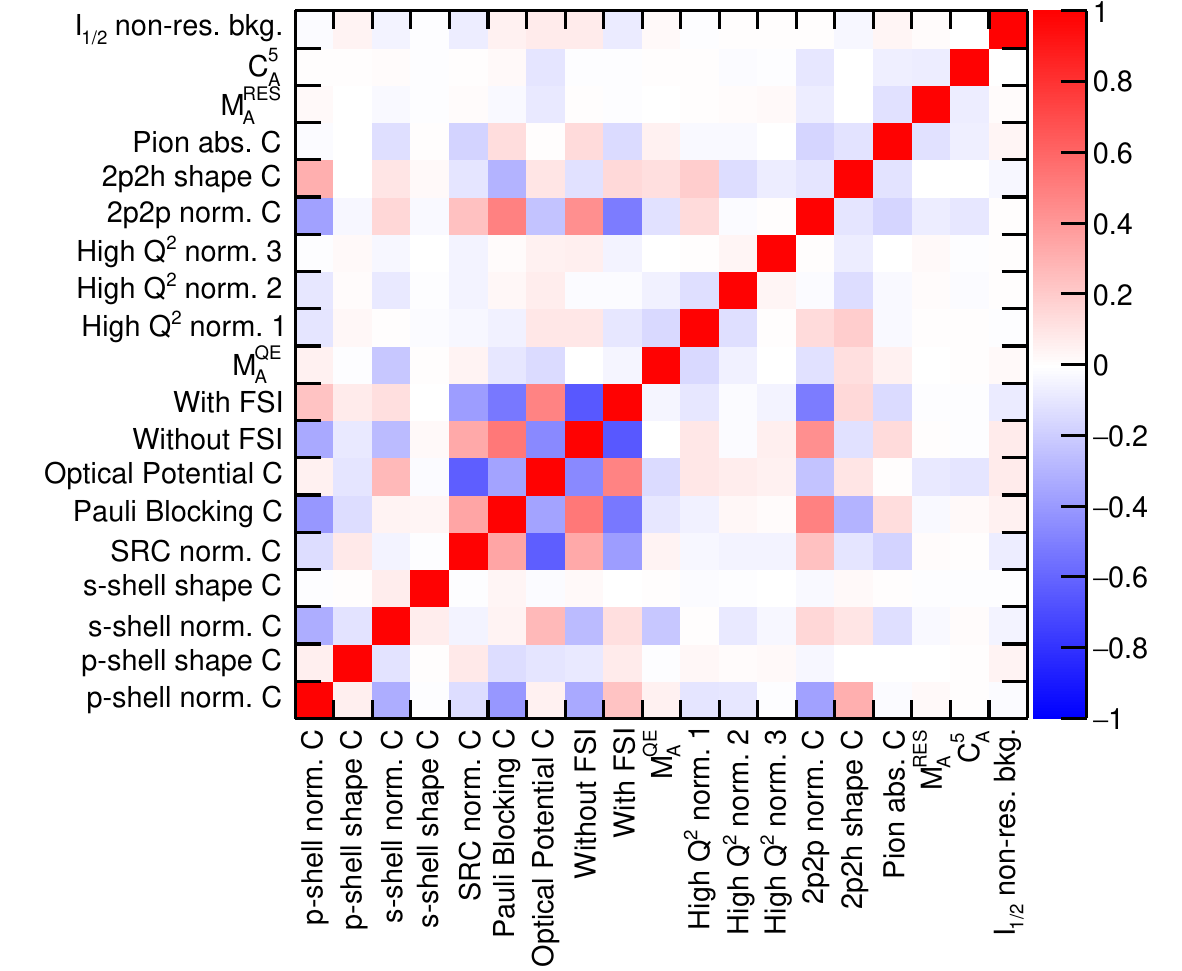}
    \caption{Postfit correlation matrices from the fit to T2K CC0$\pi0p$ measurement of lepton kinematics (top left) and CC0$\pi Np$ measurement of $\dpt$ (top right), as well as the simultaneous fit (bottom).}
    \label{fig:t2k0pNp_cov}
\end{figure*}

Figs.~\ref{fig:t2k0pNp} and~\ref{fig:t2k0pNp_dpt} show the prefit and postfit distributions from fits (a) and (b) respectively, whereas Figs.~\ref{fig:t2k0pNp_params} and \ref{fig:t2k0pNp_cov} display the prefit vs.\ postfit parameters and correlations respectively from the three fits. As previously discussed and shown in Tab.~\ref{tab:priors}, it should first be noted that the three fits do not use the same set of parameters. 

\begin{figure*}[p]
    \centering
    \foreach \a in {2,...,10}
    {
        \includegraphics[page=\a, width=0.325\textwidth]{plots/FitResults/T2K_0pNp/pmu_slices.pdf}
    }

    \caption{Prefit (red) and postfit (blue) distributions of $p_\mu$ in bins of $\cos \theta_\mu $ from fitting T2K CC0$\pi0p$ data only. The usual chi-squares as well as the number of bins are quoted in the legend. The NS chi-square $\chi^2_\mathrm{NS}$ used in the minimization is reported in Tab.~\ref{tab:chi2}.}
    \label{fig:t2k0pNp}
\end{figure*}

The binning of $\dpt$, as shown in Fig.~\ref{fig:t2k0pNp_dpt}, is rather coarse, and so even the prefit chi-squares reported in Tab.~\ref{tab:chi2} for the CC$0\pi Np$ fit are relatively low. Nevertheless, the uncertainty parametrization presented in this work yields a significantly improved postfit agreement, as demonstrated by the chi-squares. This is achieved largely thanks to the SF model shell parameters as indicated in Fig.~\ref{fig:t2k0pNp_params} (black). This fit also shows sensitivity to SRC, nucleon FSI and 2p2h shape parameters, as seen by the reduction of their postfit uncertainties. Indeed, these are effects that are probed by $\dpt$, particularly in the tail of its distribution~\cite{Lu:2015tcr}. The correlations between them that appear in the postfit correlation matrix (top left of Fig.~\ref{fig:t2k0pNp_cov}) indicate that these effects cannot be disentangled solely with $\dpt$. The T2K measurement considered here does not attempt to separate the 2p2h and FSI contributions to the tail of the distribution. There is scope to achieve a better separation by performing a measurement of $\dpt$ as a function of $\dalphat$ (another transverse variable~\cite{Lu:2015tcr}), or even as a function of outgoing lepton kinematics. Such a measurement has recently been performed by the MicroBooNE collaboration \cite{MicroBooNE:2023dwo}, and there are prospects for similar high-precision measurements using the upgraded T2K ND280 detector \cite{Abe:2019whr}.
On the other hand, even with the coarse binning, the CCQE dominance in the bulk of $\dpt$ (which, for CCQE events is just the transverse projection of the missing momentum) allows a notable constraint on the shell normalization and shape parameters. 

Contrary to the CC$0\pi Np$ fit and the oxygen + carbon analysis from Sec.~\ref{sec:fitOC}, fits including the CC$0\pi 0p$ measurement result in relatively poor chi-squares. This indicates that the model presented in this work is sufficient for analyzing results integrated over outgoing proton kinematics or binned coarsely in $\dpt$, but is insufficient to fully describe the CC$0\pi 0p$ measurement that includes only protons below 500 MeV/$c$ in its signal definition. This is likely due to insufficient freedom to alter the ratio of events with and without a proton above 500 MeV/$c$ as a function of the outgoing lepton kinematics. It is likely that the lack of detailed FSI uncertainties are at least part of the cause of this. 

Similarly to the fit of T2K data on oxygen and carbon discussed in Sec.~\ref{sec:fitOC}, the improved agreement of the model with the data in the CC$0\pi 0p$ measurement is largely driven by the increase in the PB parameter, as shown in Fig.~\ref{fig:t2k0pNp_params} (blue), which affects the forward angular region where the discrepancies are the largest. However, in this case the PB parameter remains within the prior uncertainty range. It can also be seen that OP actually moves to apply a weaker suppression of the low energy transfer region. Overall this seems to suggest that the CC$0\pi 0p$ measurement sees less of a requirement of a forward lepton angle reduction of the cross section. This might suggest that the apparent required CCQE suppression from the fit integrated over nucleon kinematics may have been acting in lieu of a need to reduce the non-QE contributions (which should be reduced in the CC$0\pi0p Np$ result) in the forward angle region. 

Fig.~\ref{fig:t2k0pNp_params} additionally shows that the $Q^2$ parameter that affects the region $0.25 \leq Q^2 < 0.50$~GeV$^2$ converges to a value significantly outside of its prior uncertainty when considering the CC$0\pi 0p$ measurement. Note that a smaller raising of the parameter was seen the carbon-oxygen measurement considered in Sec.~\ref{sec:fitOC}. Overall this might indicate a lack of freedom to vary the cross section at high momentum transfer. 


The postfit values of the parameters from the simultaneous fit of CC$0\pi 0p$ and CC$0\pi Np$ measurements are also displayed in Fig.~\ref{fig:t2k0pNp_params} (purple). Most of the parameters converge to similar values as in the CC$0\pi 0p$-only fit since the corresponding data statistically dominates the chi-square and drives the fit. In particular, the FSI parameters are less pulled by the fit in comparison with the CC$0\pi 0p$-only one thanks to the constraint from the \dpt distribution tail. The bottom panel of Fig.~\ref{fig:t2k0pNp_cov} shows strong (anti)correlations between parameters related to PB, OP, FSI and multinucleon effects. This explains the slightly different postfit values for some of these parameters between the CC$0\pi 0p$-only and the simultaneous fit. With a more statistically significant $\dpt$ measurement, we could expect a better separation between the struck nucleon-related uncertainty (FSI, SRC, 2p2h) probed by $\dpt$ and the low energy transfer effects probed mostly by the forward outgoing muon kinematics.  

\subsubsection{Fit to \minerva cross section data in CC0$\pi N$p topology on hydrocarbon}
\label{sec:fitMNV}

\begin{figure}[]
    \centering
    \includegraphics[width=0.95\linewidth]{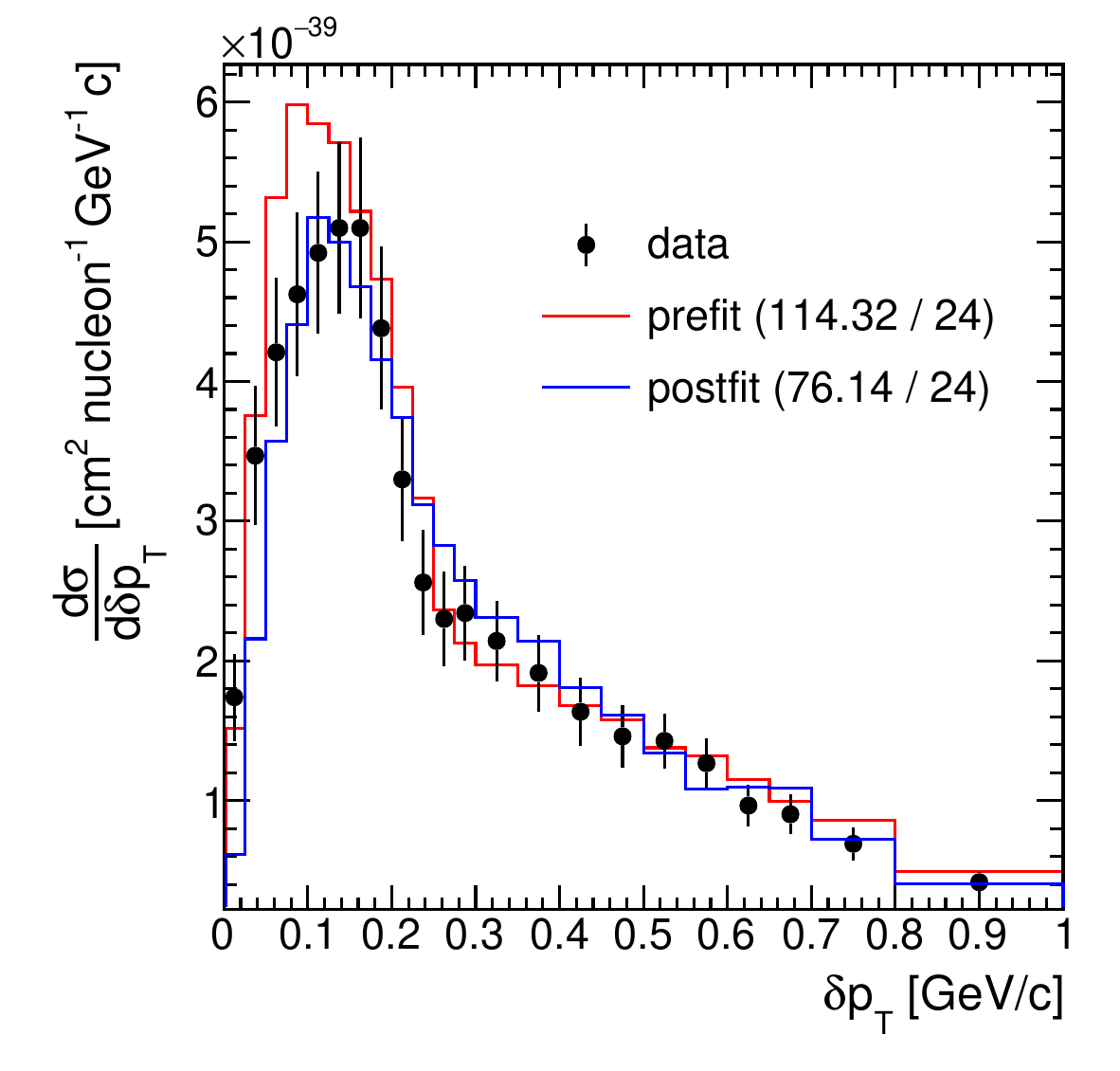}
    \caption{Prefit (red) and postfit (blue) distributions from the fit to \minerva CC0$\pi+Np$ measurement of $\dpt$ on carbon. The usual chi-squares as well as the number of bins are quoted in the legend. The NS chi-square $\chi^2_\mathrm{NS}$ used in the minimization is reported in Tab.~\ref{tab:chi2}.}
    \label{fig:mnv_dpt}
\end{figure}

\begin{figure}[]
    \centering
    \includegraphics[width=0.95\linewidth, trim={3.2cm 0 4cm 0}, clip]{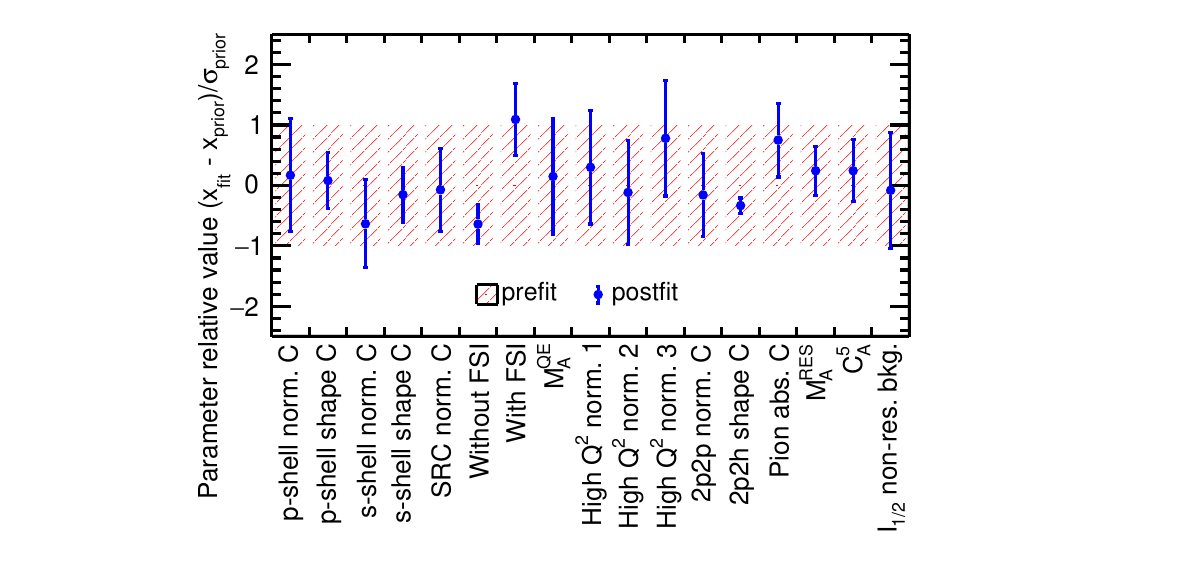}
    \caption{Prefit (red) and postfit (blue) values and constraints on the uncertainties from the fit to \minerva CC0$\pi Np$ measurement of $\dpt$ on carbon. The displayed central value for each parameter corresponds to the difference with respect to its prior value divided by the prior uncertainty as reported in Tab.~\ref{tab:priors}.}
    \label{fig:mnv_dpt_params}
\end{figure}

Results from the fit to the \minerva measurement of the CC$0\pi Np$ (considering $N$ protons above 500 MeV/$c$) cross section as a function of $\dpt$  are shown in Fig.~\ref{fig:mnv_dpt}. As suggested by the chi-square values quoted in Tab.~\ref{tab:chi2}, the data-MC prefit agreement is quite poor and is slightly improved in the postfit. Nevertheless, Fig.~\ref{fig:mnv_dpt_params} exhibits a clear sensitivity to most of the considered parameters, including the missing-momentum shape uncertainties. This is partially due to the significantly finer binning in the \minerva data in comparison with the T2K measurement. This allows a more precise probe of the nuclear effects that impact the $\dpt$ distribution. In fact, as discussed in Sec.~\ref{sec:SF}, the bulk is sensitive to Fermi motion which is mainly affected by the shell normalization and shape parameters (see Figs.~\ref{fig:shellnorm} and~\ref{fig:pshell_dpt_cthmu}). On the other hand, its tail can be altered by SRC, 2p2h, CCRES and FSI uncertainties. 

However, it is clear from the relatively high value of the postfit $\chi^2_\mathrm{NS}$ that the present parametrization of the SF systematic uncertainties does not provide enough freedom in the model to entirely cover discrepancies with the measurement. This can be attributed to the fact that, due to the higher energy of the \minerva flux, there is a significant contribution from the other interaction channels (like CCRES) through pion absorption that would need a more sophisticated parametrization. For instance, the predicted fraction of CCRES events by \texttt{NEUT} corresponds to almost $\sim 20\%$ of the CC$0\pi$ topology for \minerva, which is below $10\%$ in the case of T2K. This larger CCRES component is also responsible for the tighter constraints on the Rein--Sehgal parameters in comparison with the previous fits. The relatively poor agreement can also mean that the current parametrization of the CCQE model may need further improvements, especially for FSI effects. 

\begin{figure}[]
    \centering
    \vspace{3mm}
    \includegraphics[width=\linewidth]{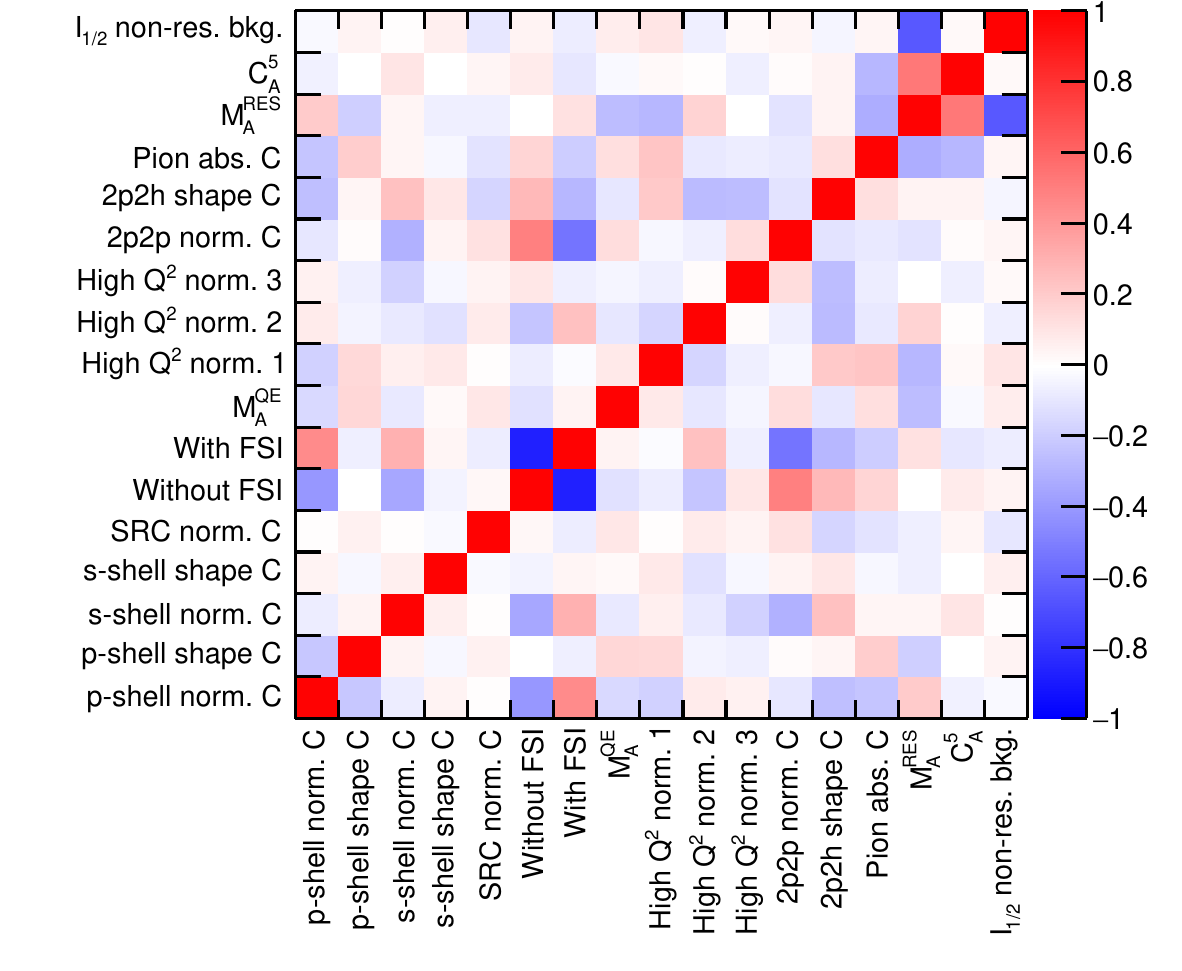}
    \caption{Postfit correlation matrix from the fit to \minerva CC0$\pi Np$ measurement of $\dpt$ on carbon.}
    \label{fig:mnv_dpt_cov}
\end{figure}

Fig.~\ref{fig:mnv_dpt_cov} reports the postfit correlation matrix for this fit. In comparison with the T2K $\dpt$ fit, we can notice that the anticorrelation between the $p$- and the $s$-shell normalization parameters is less prominent thanks to the finer binning which alleviates their degeneracy with a more precise probe of the shape of the $\dpt$ distribution. Besides, the SRC, 2p2h, FSI and CCRES uncertainties are correlated as expected since they affect the same high-$\dpt$ region.

\section{Implications for oscillation analyses}
\label{sec:impactOnOA}


In order to qualitatively evaluate the impact of this new parametrization of uncertainties for the SF model in the context of future neutrino oscillation measurements, we can consider the prefit and postfit uncertainties projected onto observables oscillation analyses are particularly sensitive to: namely the neutrino energy dependence of the cross section and the $E_\nu^\mathrm{QE}$ bias.

\begin{figure*}[]
    \centering
    \includegraphics[width=0.44\linewidth,page=1]{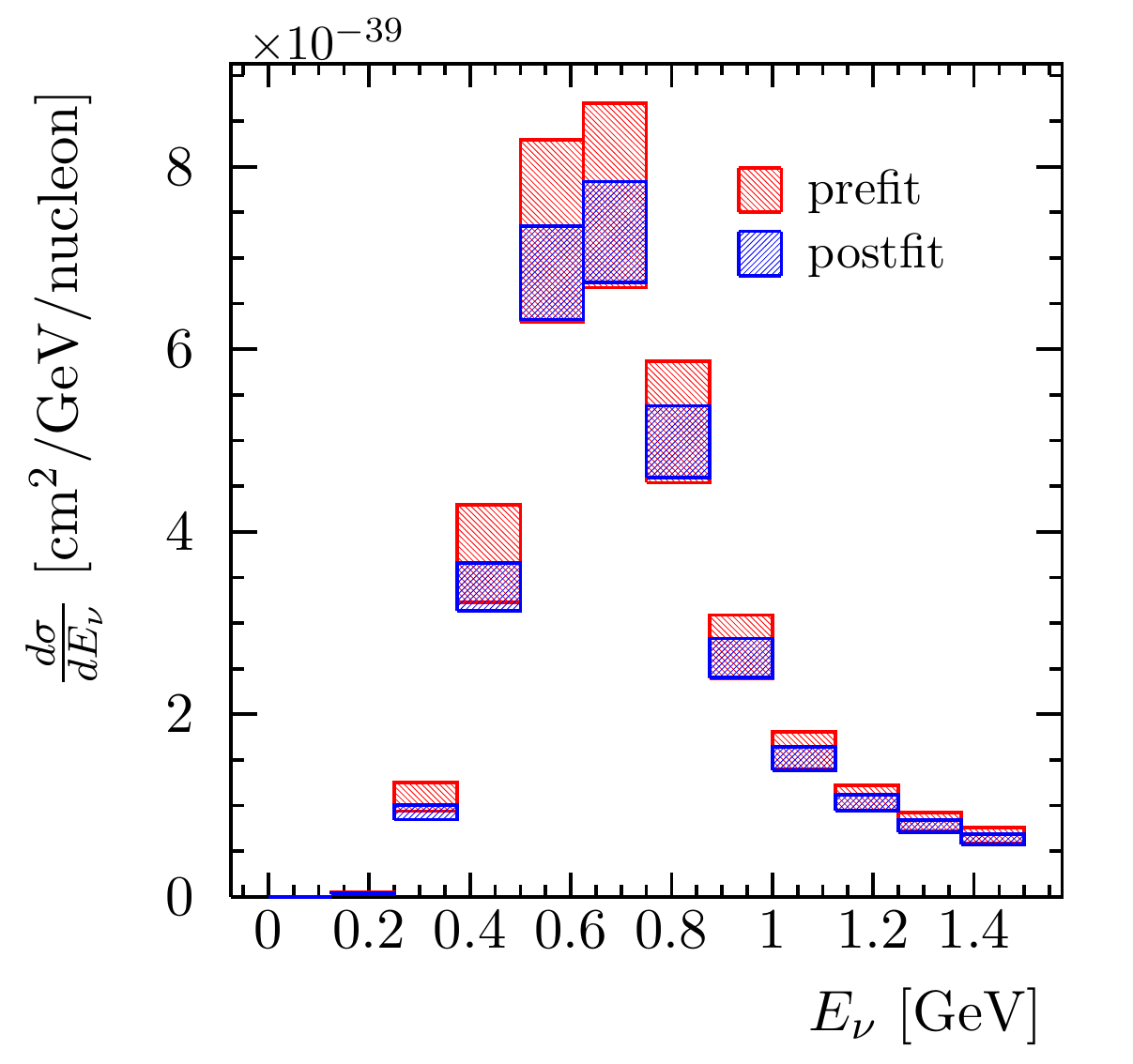}
    \includegraphics[width=0.44\linewidth,page=3]{plots/FitResults/EnubiasErrors_fixedPShellNorm.pdf}
    \includegraphics[width=0.44\linewidth,page=2]{plots/FitResults/EnubiasErrors_fixedPShellNorm.pdf}
    \includegraphics[width=0.44\linewidth,page=4]{plots/FitResults/EnubiasErrors_fixedPShellNorm.pdf}
    \caption{Prefit (red) and postfit (blue) constraints on the true neutrino energy (top) and the bias of $E_\nu^\mathrm{QE}$ (bottom) differential cross sections in the T2K beam on a carbon target following the fit to the T2K CC0$\pi$ joint measurement of lepton kinematics on carbon and oxygen (Sec.~\ref{sec:fitOC}). The plots on the left show the overall constraint on the differential cross section, whilst the plots on the right indicate the constraint on the shape of the distribution.}
    \label{fig:enu_err}
\end{figure*}

Fig.~\ref{fig:enu_err} shows the prefit and postfit spectra and constraints for the distribution of true neutrino energy and the $E_\nu^\mathrm{QE}$ bias expected in the T2K flux. These are split into showing the total constraint on a differential cross section (left) and on only its shape (right). This split is informative as it allows a separation of the overall constraint placed on the total normalization of the cross section, which should be relatively independent of the uncertainty parametrization used, from the constraint placed on the shape of the distributions.
The postfit (prefit) distributions are obtained using an ensemble of $500$ distributions sampled from the posfit (prefit) values and covariance from the fit shown in Sec.~\ref{sec:fitOC}. 

The postfit constraints on the cross section as a function of $E_\nu$ are significantly improved in comparison with the prefit, as shown in the left plots of Fig.~\ref{fig:enu_err}.
This is more visible in the bottom-left panel for the bias of $E_\nu^\mathrm{QE}$ particularly around the true $E_\nu$ (i.e.\ 0 on the plot), which is the region that is most affected by CCQE events and the SF model. The negative tail, which is more affected by multinucleon effects and CCRES interactions, is only slightly impacted since the measurement used in the fit has only a small component of these interactions.

It is clear from the bottom-right plot of Fig.~\ref{fig:enu_err} that the uncertainty model offers significant freedom in the shape of the neutrino energy bias and that this is well constrained from the fit to the T2K cross-section measurement. On the other hand, the top-right plot shows that the freedom in the shape of the neutrino energy dependence of the cross section is more limited and is not so strongly constrained by the fit.

To quantify the impact of the reduced uncertainties on the allowed space in the differential cross section as a function of the neutrino energy (top left panel of Fig.~\ref{fig:enu_err}), we can evaluate the determinant of the prefit and the postfit bin-to-bin covariance matrices of $n_b$ dimensions, where $n_b = 10$ is the number of $E_\nu$ bins. In fact, the square root of the determinant of the covariance is proportional to the volume of the $n_b$-dimensional ellipsoid that covers the $N$-sigma variations of the Gaussian errors. We find that this volume is reduced by a factor of 4.68 thanks to the constraints from the fit.



\section{Perspectives and conclusions}\label{sec:concl}

The need to use a more sophisticated parametrization of nuclear model uncertainties in neutrino oscillation experiments is becoming increasingly urgent. In this paper, we introduce a parametrization of systematic uncertainties on the inclusive and semi-inclusive predictions of the Benhar spectral function model driven mostly by the natural degrees of freedom within the model, as well as a few additional freedoms to account for the limitation of SF's PWIA description of neutrino interactions. Fits to existing T2K and \minerva CC0$\pi$ cross-section measurements show that this parametrization is able to offer a well-motivated means to improve the agreement with respect to the nominal model predictions, especially at the T2K energies. The fits were able to achieve a quantitatively good postfit agreement with T2K measurements of outgoing lepton kinematics on carbon and oxygen targets as well as for T2K measurements of $\dpt$, but the postfit agreement is less satisfactory when fitting T2K measurements of lepton kinematics within a restricted proton kinematic phase space. Overall this suggests that uncertainty parametrization is likely to be broadly sufficient for describing the inclusive CCQE cross section at T2K energies (although care is required with regards to the treatment of the parameters governing the cross section at low energy transfer), but that additional components are likely to be needed when detailed modeling of hadron kinematics is required. 

The SF model has been adopted by the T2K collaboration for neutrino oscillation analyses since 2020 and the described parametrization has been used for the latest iteration of its oscillation analysis~\cite{t2kOA2022_neutrino2022, t2koa2022wip}, which focuses on measurements of primarily outgoing lepton kinematics.
In the longer term, future improved measurements at T2K will be possible thanks to the new detectors like the Super-FGD in the upgrade of the T2K near detector~\cite{T2K:2019bbb}. Its fine-grained design will allow to precisely measure the kinematics of the hadronic products from neutrino interactions, allowing for instance the reconstruction of protons with momenta down to 300~MeV$/c$. Ref.~\cite{Dolan:2021hbw} shows quantitatively the expected improvements of the constraints on the nuclear effects described in the SF model with a simplified version of the parametrization introduced in this paper. It demonstrates how exploiting nucleon-lepton correlations can considerably constrain the uncertainties discussed in this work thanks to not only measurements of the single-transverse variables, but also an improved estimator of neutrino energy based on the sum of muon energy and nucleon kinetic energy. 

Beyond T2K, the proposed model parametrization is likely to form a reasonable starting point for other experiments that use nuclear shell models. This includes experiments that measure exclusive final states, as most of the added parameters offer a means to alter both lepton and hadron kinematics in a consistent way, although semi-inclusive cross-section measurements suggest some extensions to the parametrization, in particular related to freedoms within the FSI model, will likely be required.

\section*{Acknowledgements}
SD would like to thank Phill Rodrigues for insightful discussions concerning PPP. This work was supported by: P2IO LabEx (ANR-10-LABX-0038 – Project “BSMNu”) in the framework “Investissements d’Avenir” (ANR-11-IDEX-0003-01), managed by the Agence Nationale de la Recherche (ANR), France; the Ministry of Education and Science (2023/WK/04) and the National Science Centre (UMO-2018/30/E/ST2/00441), Poland; CNRS/IN2P3 and CEA, France; the University of Tokyo ICRR’s Inter-University Research Program FY2023 (Ref. 2023i-J-001), Japan; the Spanish Ministerio de Ciencia, Innovaci\'on y Universidades and ERDF (European Regional Development Fund) under contract PID2020-114687GB-100 and by the Junta de Andaluc\'ia grant No. FQM160, Spain; the Department of Energy, Office of Science, Office of High Energy Physics, Award DE-SC0008475; JSPS KAKENHI Grant Number JP18H05536, USA; DFG research grant (project number 517206441), Germany; the Royal Society, grant number URF{\textbackslash}R1{\textbackslash}211661, and the STFC, UK.

\bibliography{biblio}

\end{document}